\documentclass[fleqn,usenatbib]{mnras}
\usepackage{newtxtext,newtxmath}
\usepackage[T1]{fontenc}
\usepackage[utf8]{inputenc}
\usepackage{ae,aecompl}

\usepackage{graphicx}	
\usepackage{amsmath}	
\usepackage{amssymb}	

\title{Weak lensing cosmology with convolutional neural networks on noisy data}

\author[D. Ribli et al.]{
Dezső Ribli,$^{1}$\thanks{E-mail: dkrib@caesar.elte.hu (DR)}
Bálint Ármin Pataki,$^{1}$
José Manuel Zorrilla Matilla$^{2}$
\newauthor  \,\, 
Daniel Hsu$^{3},$
Zoltán Haiman$^{2}$
and István Csabai$^{1}$
\\
$^{1}$Department of Physics of Complex Systems, ELTE Eötvös Loránd University, Budapest \\
$^{2}$Department of Astronomy, Columbia University, New York, NY 10027, USA \\
$^{3}$Department of Computer Science, Columbia University, New York, NY 10027, USA%
}

\date{Accepted XXX. Received YYY; in original form ZZZ}

\pubyear{2019}

\begin{document}
\label{firstpage}
\pagerange{\pageref{firstpage}--\pageref{lastpage}}
\maketitle

\begin{abstract}

Weak gravitational lensing is one of the most promising cosmological probes of the late universe.
Several large ongoing (DES, KiDS, HSC) and planned (LSST, EUCLID, WFIRST) astronomical surveys attempt to collect even deeper and larger scale data on weak lensing.
Due to gravitational collapse, the distribution of dark matter is non-Gaussian on small scales.
However, observations are typically evaluated through the two-point correlation function of galaxy shear, which does not capture non-Gaussian features of the lensing maps.
Previous studies attempted to extract non-Gaussian information from weak lensing observations through several higher-order statistics such as the three-point correlation function, peak counts or Minkowski-functionals.
Deep convolutional neural networks (CNN) emerged in the field of computer vision with tremendous success, and they offer a new and very promising framework to extract information from 2 or 3-dimensional astronomical data sets, confirmed by recent studies on weak lensing.
We show that a CNN is able to yield significantly stricter constraints of ($\sigma_8 $,  $\Omega_m$) cosmological parameters than the power spectrum using convergence maps generated by full $N$-body simulations and ray-tracing, at angular scales and shape noise levels relevant for future observations.
In a scenario mimicking LSST or Euclid, the CNN yields 2.4-2.8 times smaller credible contours than the power spectrum, and 3.5-4.2 times smaller at noise levels corresponding to a deep space survey such as WFIRST. 
We also show that at shape noise levels achievable in future space surveys the CNN yields 1.4-2.1 times smaller contours than peak counts, a higher-order statistic capable of extracting non-Gaussian information from weak lensing maps.
\end{abstract}

\begin{keywords}
gravitational lensing: weak -- techniques: image processing -- cosmology: dark matter
\end{keywords}



\section{Introduction}

According to the standard cosmological model the small initial matter fluctuations evolved through gravitational collapse to yield the  large-scale structures in the present-day universe. This nonlinear physical process is sensitive to the model's parameters, such as the amplitude of the primordial density fluctuations ($\sigma_8$) or the total (dark and baryonic) matter content ($\Omega_m$). 
One of the central questions of modern cosmology is to recover the precise values of cosmological parameters from observations.
However, due to the complexity and non-linearity of the processes, this inversion is a nontrivial task. 

Dark Matter (DM) cannot be observed directly but its weak gravitational lensing (WL)  slightly distorts the apparent shapes of background galaxies. CFHT was the first large solid angle WL survey of several million galaxies \footnote{\url{http://www.cfhtlens.org/}}, and there are ongoing (KiDS450\footnote{\url{http://kids.strw.leidenuniv.nl/}}, DES\footnote{\url{http://www.darkenergysurvey.org}}, HSC\footnote{\url{http://hsc.mtk.nao.ac.jp/ssp/}}), and planned efforts to provide even larger, and higher resolution WL observations of over a billion galaxies (LSST\footnote{\url{https://www.lsst.org/}}, Euclid\footnote{\url{http://sci.esa.int/euclid/}}, WFIRST\footnote{\url{https://wfirst.gsfc.nasa.gov/}}).
The dark matter distribution inferred from WL measurements can then be used to constrain parameters of cosmological models through analytic models or simulations \citep{kilbinger2015cosmology}.

Modern sophisticated $N$-body and hydro-dynamical simulations \citep{pillepich2017simulating} are able to reproduce the evolution of matter distribution and with ray tracing, simulated 2-dimensional weak lensing maps can be generated \citep{vale2003simulating}. 
These simulations do not attempt to generate a particular realization of the universe that directly matches observations. Rather, compliance of the model is measured by some reduced statistical quantity that is independent of the arrangement of matter density in a particular realization.

\begin{figure}
	\includegraphics[width=\columnwidth]{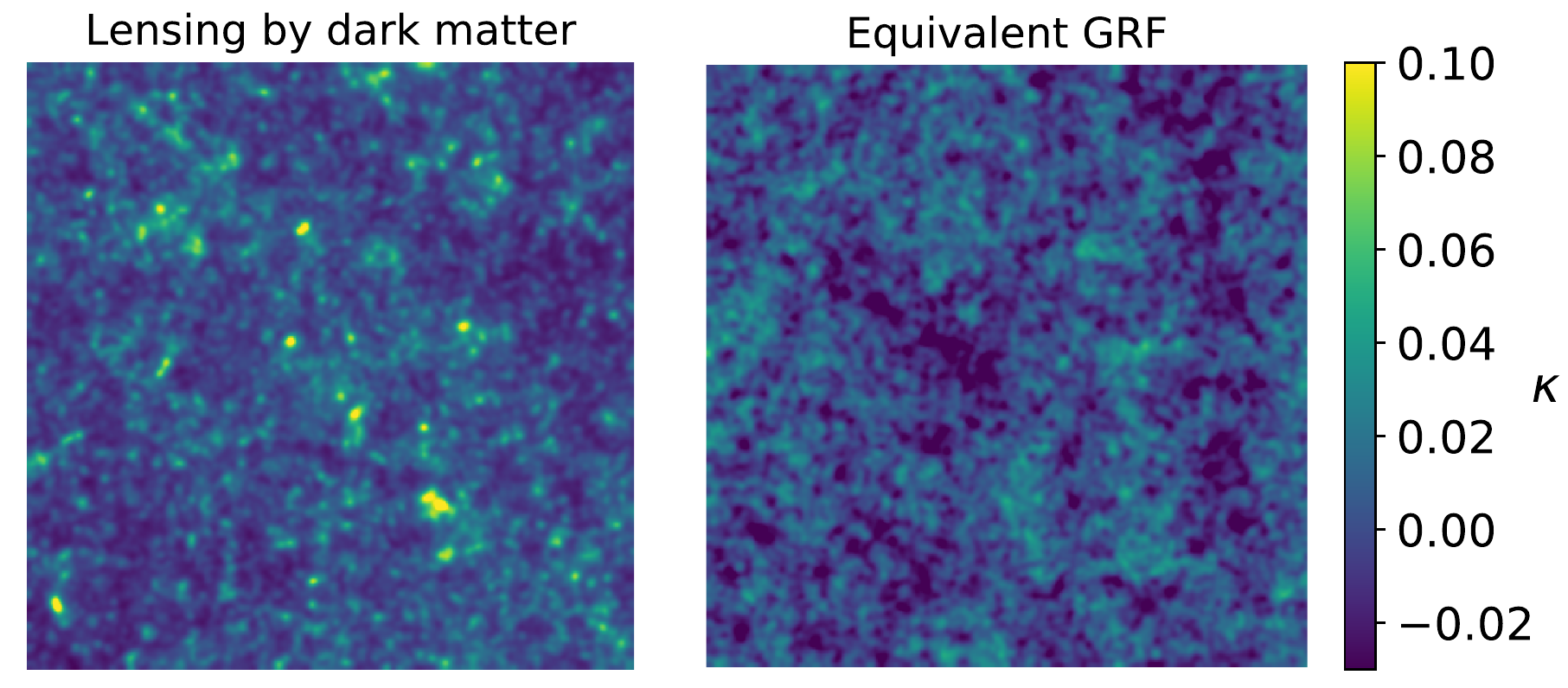}  
  \caption{A simulated convergence map (left) and its Gaussian random field equivalent (right) for cosmological parameters ($\Omega_m, \sigma_8) = (0.309,0.816)$. Extra structures on the left are clearly visible by eye.}
\label{fig:grf_comp}
 \end{figure}

Weak lensing is traditionally described using the two-point correlation function or the power spectrum of either the shear or the convergence, which fully characterize a homogeneous and isotropic Gaussian random field and do not rely on specific features of a particular individual realization.
However, on small scales, gravitational collapse distorts the Gaussian character of the initial fluctuations, as depicted on  Fig.~\ref{fig:grf_comp}, and two-point statistics are unable to capture all available information about the underlying cosmology \citep{kilbinger2015cosmology}.
Higher-order statistics~\citep{takada2003three, schneider2003, takada_jain2002, zaldarriaga2003}, peak counts~\citep{marian2009, dietrich2010cosmology, kratochvil2010probing, yang2011}, and Minkowski functionals~\citep{mecke1994, sato2001mfs, guimaraes2002mfs, kratochvil2012probing} were proposed and have been used in observations~\citep{fu2014cfhtlens,shan2014weak, liu2015cosmology, kacprzak2016cosmology, shirasaki2014statistical} to extract the remaining, non-Gaussian information from weak lensing observations.

Convolutional neural networks (CNNs) have been shown to be able to extract very complex information from images and efficiently solve nonlinear inversion problems.
Deep CNNs have revolutionized computer vision in the past six years and have became the state-of-the-art approach in virtually all computer vision tasks, such as classifying images, detecting objects or drawing pixel-wise segmentation masks.
CNNs have reached human accuracy and reduced error rates more than 10 fold in image recognition compared to previous approaches \citep{krizhevsky2012imagenet,simonyan2014very,szegedy2015going,he2016deep}.
Neural networks learn to extract information from large amounts of labeled data, without explicit feature design, which makes them attractive for various physical problems too, where hand-crafted descriptors are traditionally used to analyze data.
Convolutional neural networks capture the essence of images using sliding window filter matching operations which are, by construction, invariant to translations.
CNNs have already been used in some theoretical cosmological studies \citep{ravanbakhsh2016estimating, george2018deep, gupta2018non, ribli2018learning, fluri2018cosmological}, but their full potential remains largely unexplored.

Images of everyday objects, such as cars, cats and dogs share interesting properties with weak lensing maps, which make CNNs promising tools for cosmology.  In particular, both types of images have hidden factors which generate large variability in the potential examples with the same underlying true labels. 
The single image label ``car'' can corresponds to a vast number of different manufacturers, models, and colors photographed in different surroundings from varying viewpoints.
Weak lensing maps with the same underlying true cosmological parameters could likewise be generated with any particular initial random seed for initial matter density and velocity fields, and the weak lensing effect of the matter can be viewed from numerous different viewpoints.
For both everyday images and weak lensing maps, direct comparisons using a simple distance metric in pixel space are rendered meaningless by the large variability in potential examples with the same underlying true labels.
Cosmologists traditionally overcome this limitation using physically motivated reduced representations, such as the two-point correlation function or peak counts for direct model comparisons.
However, these representations potentially lose a large fraction of the information contained in the original weak lensing maps.
CNNs have the potential to extract additional information from the maps, independent of their particular realization, similarly to their ability to recognize cars on images, regardless of the variation of their appearance.

\subsection{Related recent work}

The subject of applying CNNs to estimate cosmological parameters from WL maps, or more broadly from large modern cosmological datasets, is in its infancy.
In an "early" attempt \citet{schmelzle2017cosmological} used deep learning in order to resolve the degeneracy of cosmological parameters from noisy data.
\cite{gupta2018non} inferred credible cosmological parameter contours from simulated noiseless convergence maps and they showed that a CNN is able to produce more accurate constraints than either the power spectrum or peak counts.
A subsequent study showed that the accuracy of a CNN on noiseless can be radically improved with a better neural network architecture \citep{ribli2018learning}.

Previous studies have also analyzed the potential of neural networks to extract cosmological information from noisy simulated convergence maps. \cite{fluri2018cosmological} used a CNN to infer cosmological parameters directly from the maps and \cite{shirasaki2018denoising} used deep learning to de-noise the maps as a pre-processing step, to perform cosmological inference by other means. 

The present study follows an approach similar to that of \cite{fluri2018cosmological}, but differs in the following aspects. First, we use a different method to simulate our training dataset. The matter density field is evolved with a full N-body code (\textsc{GADGET-2}) instead of faster, approximate methods such as \textsc{L-PICOLA} and the convergence is evaluated through full ray-tracing, instead of using the Born approximation. We note that the use of the Born approximation is adequate for estimating the power spectrum but not for higher order statistics \citep{petri2017born}. These improvements allow us to simulate the convergence field more accurately at a higher resolution (i.e. down to $1$ arcmin vs. $2.34$). As a result, the non-Gaussian information content in the training dataset is potentially higher. Second, our network architecture and training process are also different. Third, we compare the ability of the CNN to extract information with other statistics beyond the power spectrum (i.e. lensing peaks). Finally, we benchmark the performance of our CNN to that of other, previously used architectures.

\subsection{Outline of the paper}

In this work, we use weak lensing convergence maps generated with full $N$-body simulations and ray-tracing at the highest relevant angular resolution for observations (1 arcmin) in order to fully exploit non-Gaussian information.
We estimate the cosmological parameters $\Omega_m$ and $\sigma_8$ with a new CNN architecture which is able to process convergence maps covering the full $3.5 \times 3.5\deg^2$ field generated in the simulations.
We compare the predictions of the neural network to the true underlying model parameters and we invert these predictions to derive credible regions for $\Omega_m$ and $\sigma_8$.
We compare the results achieved by the CNN not only to the power spectrum, but also to peak counts.  The latter statistic has previously been found very effective in extracting non-Gaussian information, and is known to yield tighter constraints than the power spectrum in both simulations \citep{dietrich2010cosmology, kratochvil2010probing} and in real data~\citep{liu2015cosmology}.

In \S~\ref{sec:methods}, we present our methodology, briefly summarizing the simulated data suites (\S~\ref{subsec:data}), the procedure to derive constraints from the power spectrum and peak counts (\S~\ref{subsec:features}) and finally the architecture of the neural network and the constraints derived using it (\S~\ref{subsec:cnn}).
Our results are presented in \S~\ref{sec:results}.
For reference, in \S~\ref{subsec:noiseless} we first present results on noiseless convergence maps, and move on to the main result of the paper in \S \ref{subsec:noisy} where we evaluate the different methods at various noise levels relevant to ongoing and planned future observations.
We discuss our results and offer our conclusions in \S~\ref{sec:conclude}.

We also include several appendices, in which we
compare the efficiency of the map pixel size and CNN architecture introduced in this work to those used in previous studies (Appendix~\ref{subsec:archcomp}) 
and assess the impact of 
the cosmology-dependence of the covariances (\ref{subsec:varyingcov});
interpolations to unseen points in the parameter space (\ref{subsec:ninterpol}); 
confronting the CNN with an unseen realization of the initial density field (\ref{subsec:newidf}); 
augmenting the simulation suite by random transpositions and rotations (\ref{app:augmentation});
the method to split views between the training and test sets (\ref{app:splits});
and the non-uniformity of the sampling of the cosmological parameter grid around the fiducial cosmology (\ref{subsec:densegrid}).


\label{subsec:archcomp}

\section{Data and Methods}

\label{sec:methods}

\subsection{Data}
\label{subsec:data}

The dataset of images used to train and test our network is the same used in \cite{gupta2018non}. It consists of synthetic noiseless convergence maps for a suite of spatially flat $\Lambda$CDM cosmologies. Each cosmology differs in two parameters, the matter density of the present universe as a fraction of its critical density, $\Omega_m$, and the amplitude of primordial density fluctuations measured in the local universe, $\sigma_8$. These two parameters are sampled over a non-uniform 2D grid whose density increases towards a model defined by $\Omega_m=0.26$ and $\sigma_8=0.8$ (see \citealt{gupta2018non} for more details).

Each map covers a field of view of $3.5 \times 3.5 \deg^2$ with a resolution of $\approx 0.2$ arcmin ($1024 \times 1024$ pixels) and is the result of raytracing the outputs of DM-only $N$-body simulations to a redshift of $z=1$. For each cosmological model, 512 different maps were created, by building pseudo-independent past light cones from the same $N$-body simulation.  The initial matter density and velocity fields are generated with the same random seed, and the same past light cones (i.e. same viewing angles and orientations) are used in each  cosmology.  We refer the reader to \citet{gupta2018non} for a detailed description of how these data was generated.

For the present study, we further pre-processed the maps before feeding them to the network, in four steps:

\begin{enumerate}[i]
	\item We downsampled the maps by a factor of 2, speeding the training process and increasing the number of maps that fit in memory for each mini-batch.
	\item We added shape noise in the form of random white noise with a level compatible with the expected depth of upcoming galaxy surveys.
	\item We smoothed the maps with a Gaussian kernel of width equal to 1 arcmin.
	\item We applied a random horizontal and vertical flip and a random transposition during training the neural network.
\end{enumerate}

The initial downsampling does not induce a noticeable loss of information on small scales, because the pixel angular scale of the downsampled maps ($\approx 0.4$ arcmin/pixel) is still smaller than the Gaussian kernel used to smooth the maps after the addition of noise.
We validated that the initial downsampling has a negligible effect on the results achieved with the power spectrum and peak counts.

The inclusion of shape noise, resulting from the unknown intrinsic ellipticities of the galaxies whose shape is measured, brings us one step further to the application of neural networks to galaxy surveys.
For simplicity, we neglect intrinsic alignments (IA) in this paper and assume that the noise in each pixel is independent and follows a Gaussian distribution with standard deviation

\begin{equation}
\sigma_{pix} = \frac{\sigma_e}{\sqrt{2 n_g A_{pix} }},
\label{eq:shapenoise}
\end{equation}

where $\sigma_e$ is the mean intrinsic ellipticity of the galaxies in the survey (we used a value of 0.4 for this study), $A_{pix}$ is the area of each pixel, and $n_g$ is the mean galaxy density in units inverse of the ones used for the pixel area. This noise is not instrumental and can be mitigated with deeper surveys and larger $n_g$.
IA already needs to be taken into account in existing surveys (e.g. \citealt{kacprzak2016cosmology}) and its impact on our conclusions will need to be addressed in future work.

The smoothing of the maps to scales of $\approx 1$ arcmin serves two purposes. First, it filters out part of the shape noise introduced in step (ii), increasing the signal-to-noise (S/N) ratio of the data. Second, it removes information at very small scales where the presence of baryons significantly alters the matter distribution and hence the lensing signal, necessitating further modeling of baryonic physics.

Finally, the random flips and transpositions help reinforce the invariance of the network under these transformations and as a data augmentation technique, it helps alleviate the risk of over-fitting.

\subsection{Power spectrum and peak counts}
\label{subsec:features}

The power spectrum and peak counts both yield a fixed descriptor for each map, hence they are treated in a unified framework.

The power spectra of the convergence maps are measured using the LensTools \citep{petri2016mocking} python package in 20 logarithmic bins in spherical harmonic index between $100 \leq \ell \leq 3.75\times 10^4$.
This range covers the angular scales present in the unsmoothed maps. While smoothing suppresses power at $\ell \lessapprox 10^4$, we chose a set of bin edges that would allow for direct comparisons with unsmoothed maps.

Peaks are defined as local maxima on a map, and "peak counts" refer to the binned histogram of a set of peaks as a function of their height. It has been shown that 5-10 bins are sufficient to capture the cosmological information in single-redshift analyses of peak counts \citep{petri2016tomography}. Here we chose 20 bins to limit the bias correction in the precision matrix. The minimum and maximum values were chosen in units of the mean (noiseless) r.m.s. $\kappa$ to avoid empty bins or a singular precision matrix in the models used for its estimation. Finally, since the dynamic range on each map is limited ($\kappa_{\rm min}\approx-0.03$ and $\kappa_{\rm max}\approx 0.19$), the bins were spaced linearly.

Parameter confidence contours are calculated in the same fashion for peak counts and the power spectrum analysis, with a standard Gaussian likelihood analysis. The probability of a cosmological parameter given a mock observation map can be expressed with likelihoods using Bayes' theorem,
\begin{equation}
P(\theta | \boldsymbol{d} ) = \frac{ P( \boldsymbol{d}| \theta ) \, P(\theta) }{P(\boldsymbol{d})}.
\label{eq:bayes}
\end{equation}
Here $\theta$ represents the cosmological parameters, in this case: $\Omega_m, \sigma_8$, and $\boldsymbol{d}$ denotes the descriptors measured in a mock observation. We adopt a flat prior equivalent to the convex hull of the simulation grid, and the denominator is simply a normalizing factor if we assume that $\Lambda$CDM is the true underlying cosmological model.

For the likelihood function, we chose a multivariate Gaussian distribution with a constant determinant in the denominator, 
\begin{equation}
 P( \boldsymbol{d} | \theta ) \propto \exp \left( -\frac{1}{2}  [\boldsymbol{d - \mu} (\theta)] \widehat{C^{-1}(\theta)} [\boldsymbol{d - \mu }(\theta)]   \right)
\end{equation}
which we refer to as "semi-varying" covariance. This semi-varying covariance scheme was found to be sufficiently accurate for peak counts in a previous study \citep{matilla2016dark}. Here $\mu(\theta)$ denotes the average measured descriptors, and $\widehat{C(\theta)}$ their covariance for a given cosmology.

We estimate the mean values and the unbiased covariance matrices of the descriptors at each $(\Omega_m, \sigma_8)$ point on the simulation grid using all 512 convergence maps for each cosmology.
The numerically estimated covariance matrices are de-biased following \cite{dietrich2010cosmology} as
\begin{equation}
\widehat{C^{-1}(\theta)} = \frac{N-d-2}{N-1} \widehat{C^{-1}(\theta)},
\end{equation}
where $N=512$ is the number of maps per cosmology and $d=20$ is the dimension of either observable (the power spectrum or peak counts).

The mean descriptors and their covariances are calculated using linear interpolation on a regular grid with (300$\times$300) points in the $\Omega_m$ interval $[0.18, 0.7]$ and $\sigma_8$ interval $[0.2, 1.2]$.

The likelihood values for a mock observation are calculated at each new regular grid point using the interpolated mean descriptors and the interpolated covariances in the exponent.

The likelihoods are normalized to integrate to unity, and contours containing 95\% or 68\% of the estimated total probability are defined as credible parameter regions. We define the figure of merit as the inverse of the area of the 68\% confidence region. All credible contours presented in this work are shown for a single mock observation covering a simulated $3.5 \times 3.5 \deg^2$ field.

\subsection{Convolutional neural network}
\label{subsec:cnn}

Our convolutional neural network (CNN) maps $512\times512$ pixel sized convergence maps into two numbers. We train it so that the output corresponds to the cosmological parameters used to generate the maps, $\Omega_m$ and $\sigma_8$.
The network shares its overall architecture with successful image classifiers \citep{krizhevsky2012imagenet, simonyan2014very, szegedy2015going, redmon2017yolo9000}.

The CNN consists of 18 convolutional layers, its neurons using rectified linear units (ReLU) as activation function. ReLU units introduce the needed non-linearities at a small computational cost (their gradient is constant, zero for a negative input and positive otherwise).
Each convolutional layer, except for the last one, is followed by a batch normalization layer.
Batch normalization rescales the activations prior to each optimization step, similar to the whitening typically applied to the network's input. This generally speeds up the training and offers some regularization \citep{ioffe2015batch, shibani2018batchnorm}.

We do not pad the activation maps before each convolution with zeros. These activation maps are downsampled using average pooling, allowing the network to learn correlations at increasing angular scales with a fixed convolution kernel size while speeding training (through reduced inner representations of the activations).

After the convolutional section, a single dense layer with two linear units outputs the predictions for cosmological parameters ($\Omega_m, \sigma_8$). This represents a simplification compared with the architectures used in past applications to cosmology \citep{gupta2018non,fluri2018cosmological,ribli2018learning} and follows recent developments in the field of image recognition, where multiple dense layers (with dropout) have been found to be expendable \citep{szegedy2016rethinking, he2016deep, xie2017aggregated, huang2017densely}. A detailed schema of the network is presented in Table [\ref{tab:nn}].

\begin{table}
\centering
\begin{tabular}{r|l|r}
\# & Layers & Output size \\\hline
1 & Convolution ($3 \times 3 $)    & $510 \times 510 \times 32 $ \\
2 & Convolution ($3 \times 3 $)    & $508 \times 508 \times 32 $ \\
- & Average Pooling ($2 \times 2$)  & $254 \times 254 \times 32 $ \\
3 & Convolution ($3 \times 3 $)    & $252 \times 252 \times 64 $ \\
4 & Convolution ($3 \times 3 $)    & $250 \times 250  \times 64 $ \\
- & Average Pooling ($2 \times 2$)  & $125 \times 125  \times 64 $ \\
5 & Convolution ($3 \times 3 $)    & $123 \times 123  \times 128 $ \\
6 & Convolution ($1 \times 1 $)    & $123 \times 123 \times 64 $ \\
7 & Convolution ($3 \times 3 $)    & $121 \times 121 \times 128 $ \\
- & Average Pooling ($2 \times 2$)  & $60 \times 60  \times 128 $ \\
8 & Convolution ($3 \times 3 $)    & $58 \times 58 \times 256 $ \\
9 & Convolution ($1 \times 1 $)    & $58 \times 58 \times 128 $ \\
10 & Convolution ($3 \times 3 $)    & $56 \times 56 \times 256 $ \\
- & Average Pooling ($2 \times 2$)  & $28 \times 28 \times 256 $ \\
11 & Convolution ($3 \times 3 $)    & $26 \times 26 \times 512 $ \\
12 & Convolution ($1 \times 1 $)    & $26 \times 26 \times 256 $ \\
13 & Convolution ($3 \times 3 $)    & $24 \times 24  \times 512 $ \\
- & Average Pooling ($2 \times 2$) &  $12 \times 12 \times 512 $ \\
14 & Convolution ($3 \times 3 $)    & $10 \times 10 \times 512 $ \\
15 & Convolution ($1 \times 1 $)    & $10 \times 10 \times 256 $ \\
16 & Convolution ($3 \times 3 $)    & $8 \times  8  \times 512 $ \\
17 & Convolution ($1 \times 1 $)    & $8 \times  8  \times 256 $ \\
18 & Convolution ($3 \times 3 $)    & $6 \times  6  \times 512 $ \\
 & Average Pooling ($ \times $) & $1 \times 1 \times 512 $ \\
19 & Dense  & 2 \\
\end{tabular}
\caption{Neural network architecture. Number of trainable parameters: 8,327,586 . }
\label{tab:nn}
\end{table}

\begin{figure}
    \includegraphics[width=\columnwidth]{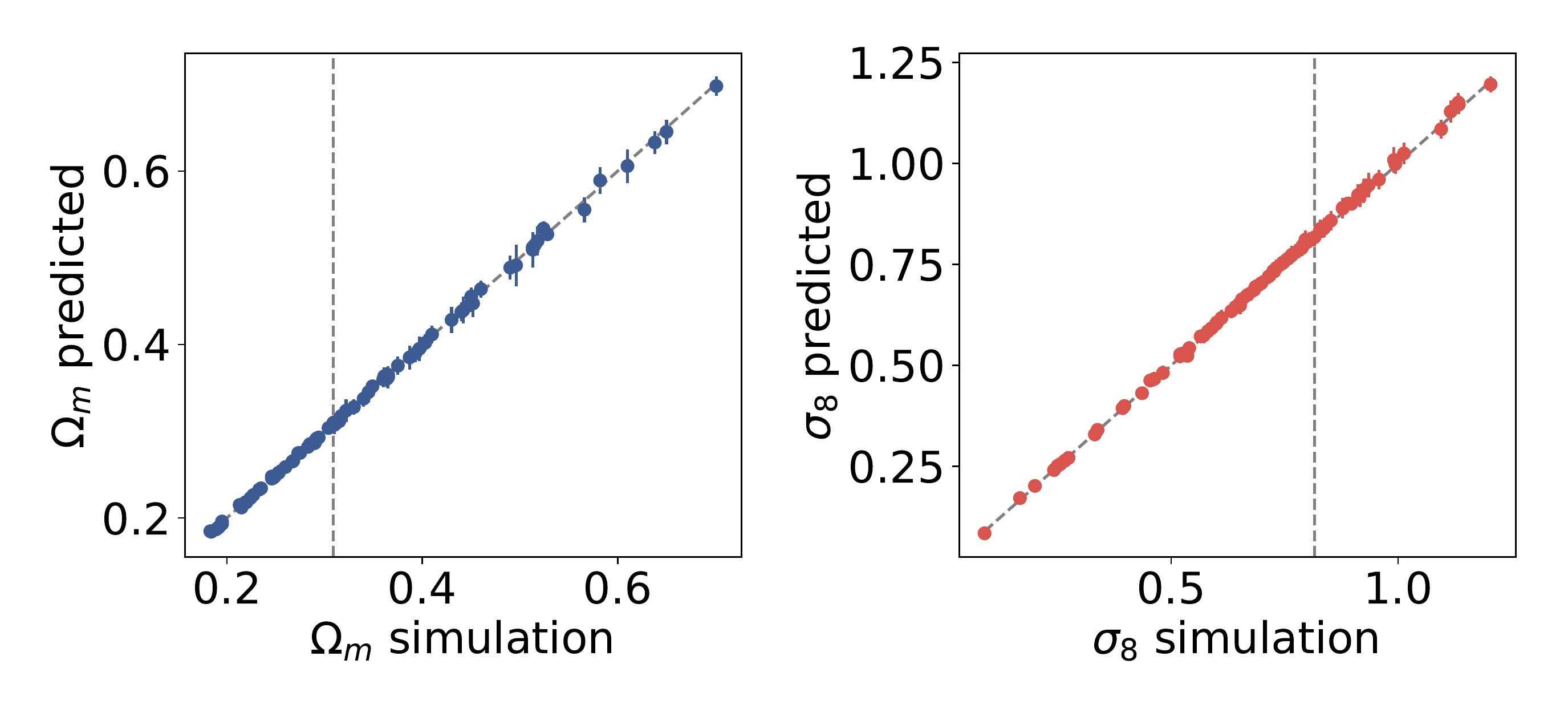}
  \caption{The CNN predicts cosmological parameters accurately on unseen noiseless convergence maps in the set. 
  The plots show the true parameters used in the simulation and the values predicted from by the neural network.
  Dots mark the mean prediction for the 153 convergence maps in the test set for a given pair of parameters $(\Omega_m, \sigma_8)$. 
  The error bars represent the standard deviation of predictions over these 153 these maps. The size of the error bars is generally too small to be visible.
}
\label{fig:scatter_nless}

      \includegraphics[width=\columnwidth]{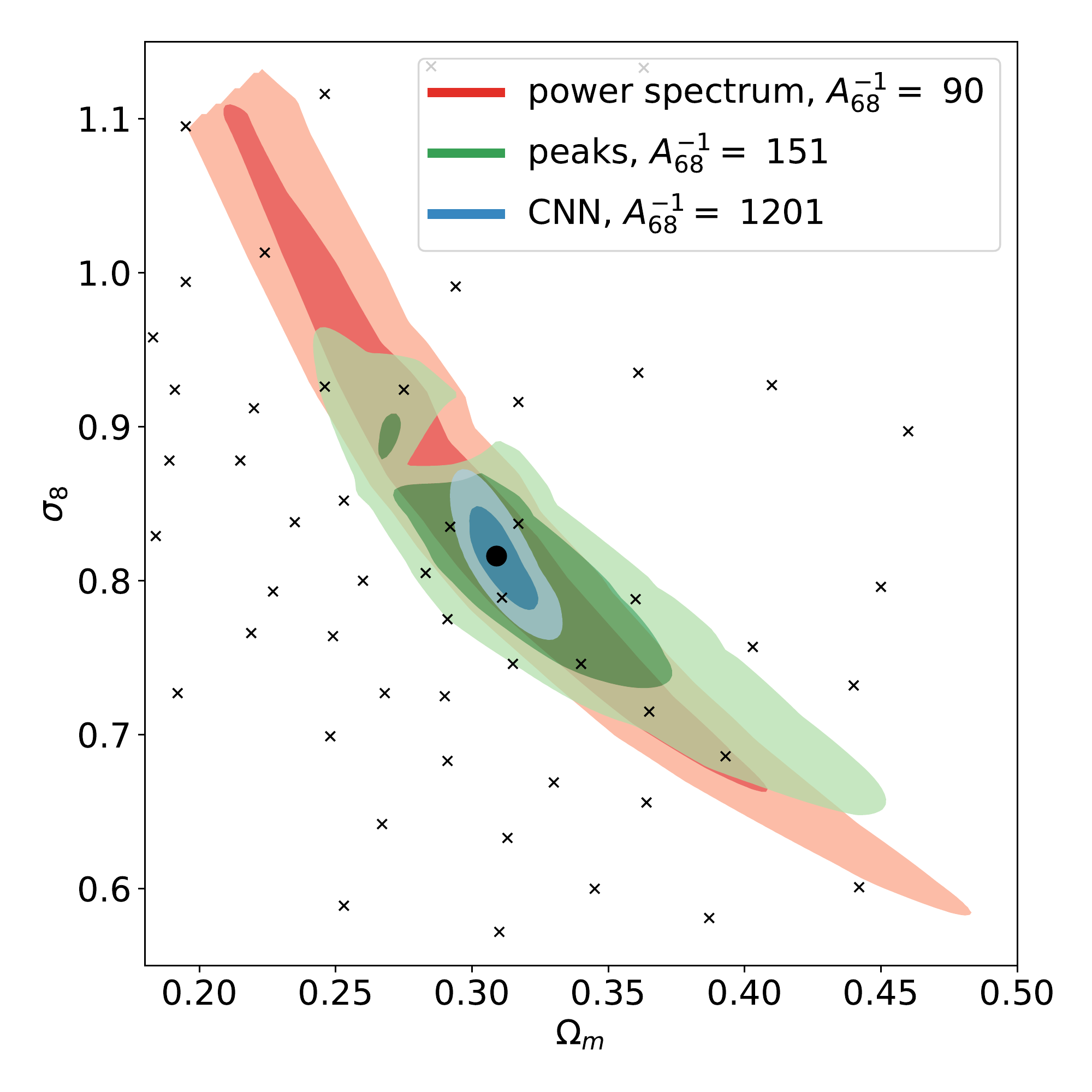}
    \caption{ 
    The CNN yields significantly smaller credible cosmological parameter contours than the power spectrum or peak counts on noiseless convergence maps.
    The mock observation is assumed to coincide with $(\Omega_m=0.309, \sigma_8=0.816)$.
    The lighter and darker regions show 95\% and 68\% confidence areas, respectively. The black crosses represent the parameter pairs of our underlying simulation suite and the larger black dot marks the mock observation.}
    \label{fig:contours_nless}
\end{figure}

\begin{figure*} 
    \includegraphics[width=1.8\columnwidth]{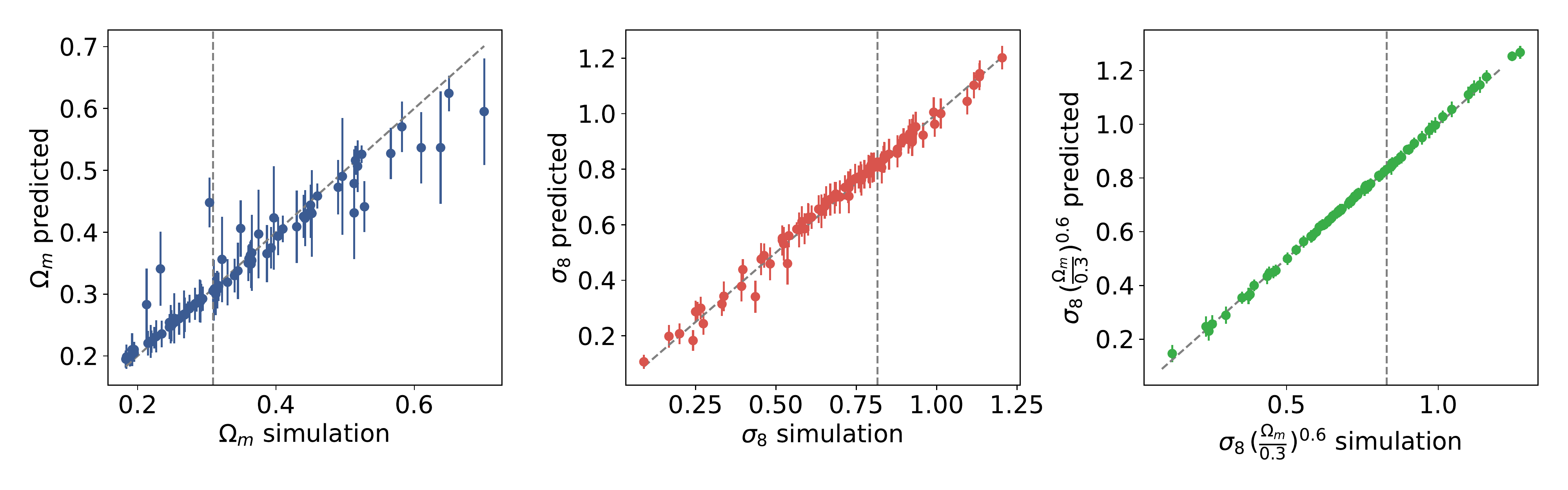} 
    \includegraphics[width=1.8\columnwidth]{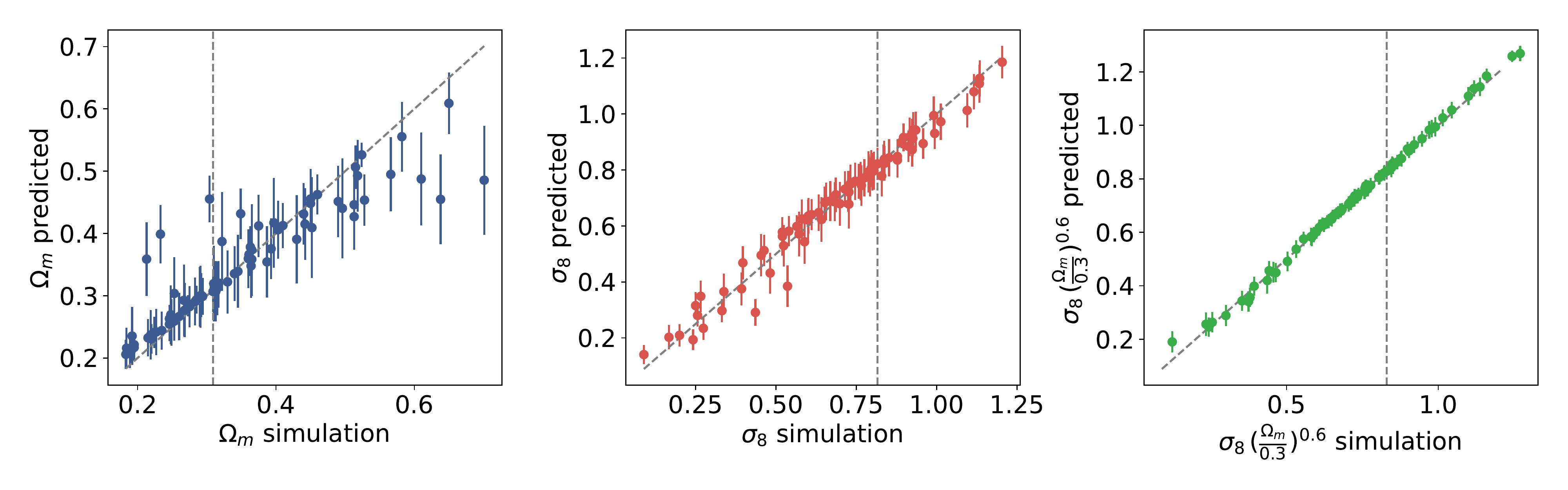}  
  \caption{Predictions of the CNN in the presence of shape noise.
  Each panel shows the original parameters used in the simulation and the values predicted from the lensing maps by the neural network.
  Dots mark the mean of the predictions for the 153 convergence maps in the test for a given pair of cosmological parameters $(\Omega_m, \sigma_8)$. 
  The error bars represent the standard deviation of predictions on these maps.
  \textbf{Top}:  Shape noise with 75 galaxies arcmin$^{-2}$.
  \textbf{Bottom}: Shape noise with 30 galaxies arcmin$^{-2}$.
}
\label{fig:scatter}
\end{figure*}

The loss function of our neural network is defined as the mean absolute difference of the predicted and true underlying $\Omega_m $ and $\sigma_8$ parameters, or mean absolute error (MAE).
Note that when we report MAE values we take the mean of the $\Omega_m $ and $\sigma_8$ errors.

The neural network is trained for 30 epochs with stochastic gradient descent, in mini-batches of 32 maps. 
The initial learning rate is 0.005 and the learning rate is divided by $10$ after epochs (10,15,20,25). During training, the maps are augmented with random horizontal and vertical flips and random transpositions, yielding a total
of $2\times 2\times 2=8$ possible combinations.
During testing each image is evaluated with all 8 of these combinations, and the predictions made on these augmented maps are averaged in order to obtain a single final prediction.  A new shape noise realization is generated on the fly for each epoch during training in order to avoid overfitting the noise.

We split the realizations into a training set (70\%) and a test set (30\%) based on the specific past light cones used to ray-trace the convergence maps.
We use a predefined fixed learning rate schedule for all experiments, and we do not use early stopping, therefore a validation split is not needed.
It is important to emphasize that we do not split the maps randomly:  each specific past light cone can only be found in either the test set or the training set.
This corrects a potential problem with random train-test splits which arises from using the same seed for each initial density field. This issue was not noticed in previous work \citep{gupta2018non,ribli2018learning}; a detailed discussion of the issue can be found in Appendix~[\ref{app:splits}]. \\

Gaussian likelihood analysis is done very similarly to the case of peak counts and the power spectrum.
The main difference is that the observed data $\boldsymbol{d}$ in equation \eqref{eq:bayes}, which is the peak count histogram or the power spectrum for the fixed-descriptor methods, is replaced by the predictions of the neural network for ($\Omega_m, \sigma_8$).
We estimate the mean and the unbiased covariance of the predictions of the neural network on the test data at each point on the simulation grid, which consists of the 153 convergence maps per cosmology in the test set.
The number of observables in the case of the CNN is therefore $d=2$.  For the CNN we do not assume constant determinant in the multivariate Gaussian distribution, therefore we evaluate likelihoods using interpolated covariances both in the exponent and in the denominator, 
\begin{equation}
 P( \boldsymbol{d} | \theta ) \sim  \frac{1}{\sqrt{|\widehat{C}|} }  \exp \left( \frac{-1}{2}  [\boldsymbol{d - \mu} (\theta)] \widehat{C^{-1}(\theta)} [\boldsymbol{d - \mu }(\theta)]   \right).
\end{equation}

The source code used for training the CNN and evaluating peak counts and the power spectrum and producing the figures and results in the this paper is available online \footnote{\url{https://github.com/riblidezso/weaklensingCNN}}.


\section{Results}
\label{sec:results}

\subsection{Noiseless maps at 1' angular resolution}
\label{subsec:noiseless}



The main focus of the current work is convergence maps with additional shape noise, however, for reference, we trained our neural network on noiseless convergence maps first.
The predictions of the CNN on the unseen convergence maps from the test set are very accurate [Fig. \ref{fig:scatter_nless}].
We derive credible contours of cosmological parameters for a mock observation with $(\Omega_m=0.309, \sigma_8=0.816)$ using the CNN, the power spectrum and peak counts.
The mock observational data for the power spectrum and peak counts is the mean descriptor calculated using all the $512$ convergence maps from the given cosmology.
In the case of the CNN, the mock ``observational data'' are the mean predictions for the cosmological parameters $(\Omega_m, \sigma_8)$ on the $153$ unseen convergence maps in the test set from that cosmology.
The contours derived from each model are shown in Fig.~[\ref{fig:contours_nless}].
On noiseless maps at 1' angular resolution, the CNN produces approximately $8\times$ smaller 68\% confidence areas than peak counts, and more than $13 \times$ smaller than the power spectrum.
The fine details of the CNN also matter; the architecture presented in this work is significantly more accurate than the one used previously in \citet[][see direct comparisons in Appendix~\ref{subsec:archcomp}]{gupta2018non}.

The results show that there is additional information in noiseless mock convergence maps accessible to the CNN at observationally relevant angular scales.
In a realistic scenario, at least part of that additional information may not be accessible due to the presence of noise. Even in the absence of atmospheric and instrumental effects, the ellipticity measured on a single galaxy is mainly due to its shape or inclination, not lensing. Averaging measurements over nearby galaxies takes advantage of the correlation of the lensing signal in nearby regions of the sky, but local tidal fields can align nearby galaxies further complicating the extraction of the lensing signal. Even if the effect of intrinsic alignment can be mitigated with priors informed from spectroscopic and radio polarization measurements \citep{blain2002detecting,morales2006technique,huff2013cosmic,kilbinger2015cosmology} and deeper surveys will improve the statistics at any given angular resolution, we cannot expect future surveys to be noise-free, and it is necessary to analyze the performance of neural networks in the presence of noise.

\begin{figure}
    \includegraphics[width=\columnwidth]{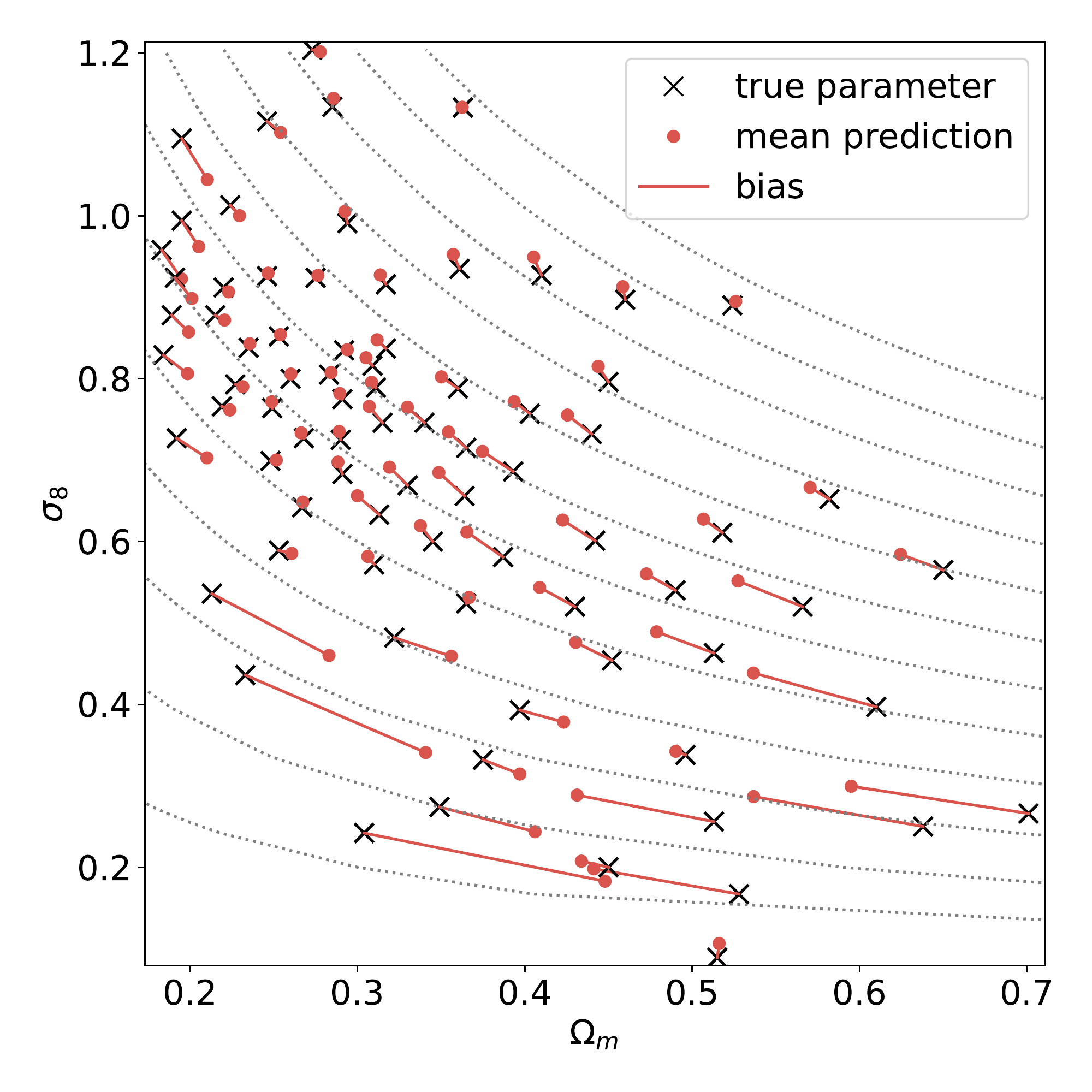}
    \caption{ Predictions are systematically biased toward the center of the simulation grid in the presence of shape noise.
    Bias is most prominent along the $\sigma_8-\Omega_m$ degeneracy (constant $\Sigma_8$).
    The black crosses represent the true cosmological parameters used in the simulations, the red dots show the corresponding CNN predictions, and the red lines mark the bias of these predictions.
    Grey dotted lines correspond to constant values of $\Sigma_8$.}
    \label{fig:bias}
\end{figure}

\begin{figure*}
\includegraphics[width=2\columnwidth]{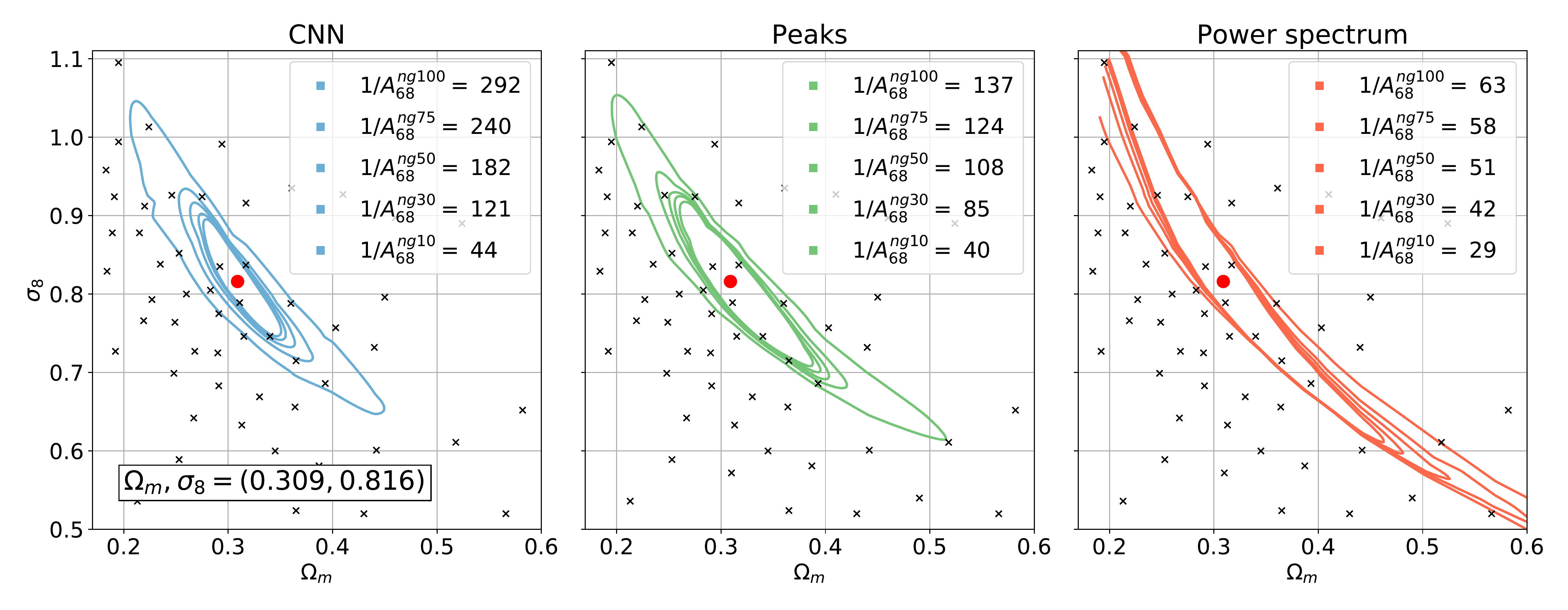}
\includegraphics[width=2\columnwidth]{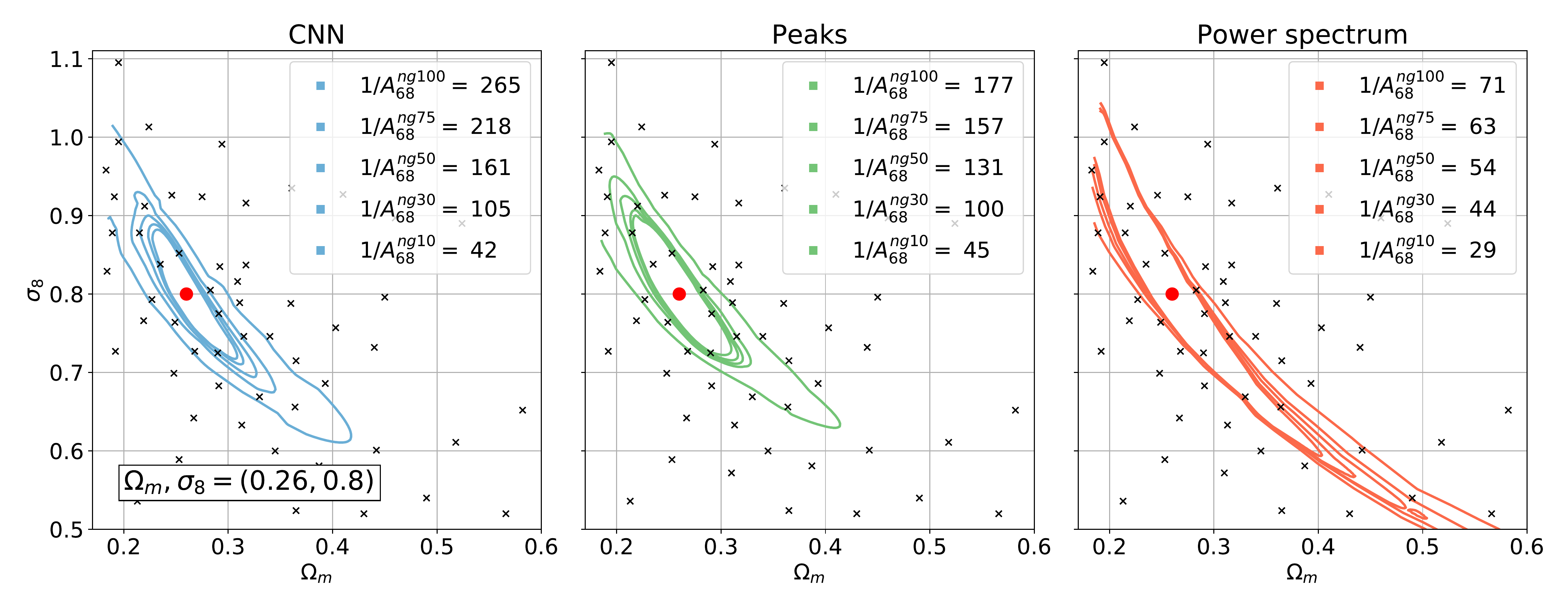}
    \caption{ 
    The CNN is able to reduce the credible areas of cosmological parameters compared to the power spectrum or peak counts, even in the presence of various levels of shape noise.
    The lines show 68\% confidence areas. The black crosses represent the parameters of our simulation grid.
    The red dot marks the cosmology of the mock observations. 
    The top panel shows a mock observation with ($\Omega_m , \sigma_8$)=(0.309,0.816).
    The bottom panel shows a mock observation with ($\Omega_m , \sigma_8$)=(0.26,0.8).}
    \label{fig:contours}
\end{figure*}

\subsection{ 1' angular resolution maps with shape noise}
\label{subsec:noisy}

As a next step towards the application of neural networks on observed convergence maps, we train a CNN on 1' per pixel angular resolution convergence maps with additional shape noise. 
We explore 5 scenarios for noise levels, corresponding to 10, 30, 50, 75 and 100 galaxies arcmin$^{-2}$.
A noise level with 10 galaxies arcmin$^{-2}$ describes typical ground-based surveys such as CFHTLens, DES or KiDS.
30 galaxies are approximately the noise level targeted by LSST or Euclid, 50-75 galaxies may be accessible in future space surveys such as WFIRST.
The 100 galaxies arcmin$^{-2}$ case is an optimistic scenario for a future-generation space-based survey.

The predictions of the neural network are shown in
Fig.~\ref{fig:scatter}. They are less accurate in the presence of shape noise: they show significant scatter, and the highest-$\Omega_m$ cosmologies are predicted with a systematic bias.
The origin of this systematic bias is not completely evident from the one dimensional scatter plots.
However, a two-dimensional diagram of the mean errors of the predictions, shown in Fig.~\ref{fig:bias}, clearly reveals that predictions are systematically and consistently biased towards the central values of ($\Omega_m , \sigma_8$) on the simulation grid. This bias occurs along the $\Sigma_8$ degeneracy, where $\Sigma_8 \equiv \sigma_8 \left(\frac{\Omega_m}{0.3}\right)^{0.6}$.
It turns out that apparent outliers on the one dimensional scatter plots are simply cosmologies which have a strong systematic bias.

The reason for bias towards the center of the data distribution is understandable in a hard regression problem. 
If the model is not able to correctly predict the target values then the median is a "good prediction" as it minimizes the mean absolute error loss function, when no further information is available.

The bias is especially strong along the degeneracy between $\sigma_8$ and $\Omega_m$, because there is not enough information to resolve the degeneracy, however, there is little bias perpendicular to the degeneracy because there is enough information to predict $\Sigma_8$ relatively well.
As pointed out previously \citep{gupta2018non,fluri2018cosmological}, Gaussian likelihood analysis handles biased predictions correctly, because this bias is converted into variance during the calculation of credible parameter contours.
Therefore while the predictions are systematically biased, the calculated credible parameter contours remain unbiased.

We infer the credible contours of cosmological parameters for the CNN, peak counts, and the power spectrum at each noise level for two mock observations with different $\sigma_8$ and $\Omega_m$ values [Fig. \ref{fig:contours}].
The first mock observation has cosmological parameter obtained by Planck collaboration \cite{ade2016planck} ($\Omega_m , \sigma_8$)=(0.309,0.816) and the other has both smaller $\Omega_m $ and  $\sigma_8$ values (0.26,0.8). 

As the main result of the current study, we show that the CNN yields 3.7-4.6 times smaller contours than the power spectrum at a noise level of 100 galaxies arcmin$^{-2}$, and 3.5-4.1 times smaller at 75 galaxies arcmin$^{-2}$, demonstrating a large advantage for future space-based surveys.
The CNN also shows a significant advantage in a scenario achievable in LSST or Euclid, with 2.4-2.8 times smaller contours than achievable with the power spectrum.
We show for the first time that in the presence of shape noise the CNN also outperform peak counts, which is an effective method specifically designed to capture the signs of non-Gaussian densities in convergence maps.
For noise levels achievable in future space surveys the CNN produces 1.5-2.1 times smaller contours than peak counts at 100 galaxies, and 1.4-1.9 smaller at 75 galaxies arcmin$^{-2}$.
At a noise level corresponding to LSST or Euclid, the relative advantage decreases to 1.05-1.42.

Interestingly, the CNN and peak counts behave differently at the two mock observations. At cosmology ($\Omega_m , \sigma_8$)=(0.26,0.8) the CNN is less accurate than at the Planck parameters, while peak counts turn out to more accurate.
Therefore we find higher improvements achieved by the CNN over peak counts at ($\Omega_m , \sigma_8$)=(0.309,0.816) than in the mock observation with smaller cosmological parameters, ($\Omega_m , \sigma_8$)=(0.26,0.8).

\begin{table*}
\centering
\begin{tabular}{l|r|r|r|r|r|r}
$A_{68}$ ratio & Noiseless & $100\, \frac{\rm gal}{\rm arcmin^2}$  & $75\, \frac{\rm gal}{\rm arcmin^2}$  & $50\, \frac{\rm gal}{\rm arcmin^2}$  & $30\, \frac{\rm gal}{\rm arcmin^2}$ & $10\, \frac{\rm gal}{\rm arcmin^2}$  \\\hline
 \hline
Power spectrum / CNN & $13$ & $3.7-4.6$ & $3.5-4.1$ & $3-3.6$ & $2.4-2.8$ & $1.4-1.5$ \\
Peak counts / CNN & $8$ &   $1.5-2.1$ & $1.4-1.9$ & $1.2-1.7$ & $1.05-1.42$ & $0.9-1.1$\\
\hline

\end{tabular}
\caption{The table lists the relative sizes of the 68\% credible contour areas of the power spectrum and peak counts compared to the CNN. The CNN achieves smaller 68\% credible contour areas than the power spectrum for any noise level, and also outperforms the peak counts when the galaxy density is at least $30 / {\rm arcmin^2}$. }
\label{tab:improvements} 
\end{table*}


\section{Discussion}
\label{sec:conclude}

\subsection{Further experiments}

The neural network in this study is different from the ones used in previous works \citep{gupta2018non, ribli2018learning, fluri2018cosmological}. It does not use a tiling scheme that cuts the input maps into smaller regions; it is instead fed full 512$\times$512-pixel maps as its input. The larger inputs make it easier for the network to learn features on larger angular scales. Its architecture includes batch normalization layers (except for the last layer), and its implementation proved critical, allowing us to train networks on noisy data without the need to dial-in the noise level on networks pre-trained on noiseless data as in \citet{fluri2018cosmological}.
The use of drop-out layers for regularization as in \citet{gupta2018non} also became unnecessary. When compared with the architectures used in the above-mentioned previous studies, on the same dataset, the architecture used in this study proved to be more accurate (see Appendix~\ref{subsec:archcomp} for direct comparisons).

\begin{figure}
    \includegraphics[width=\columnwidth]{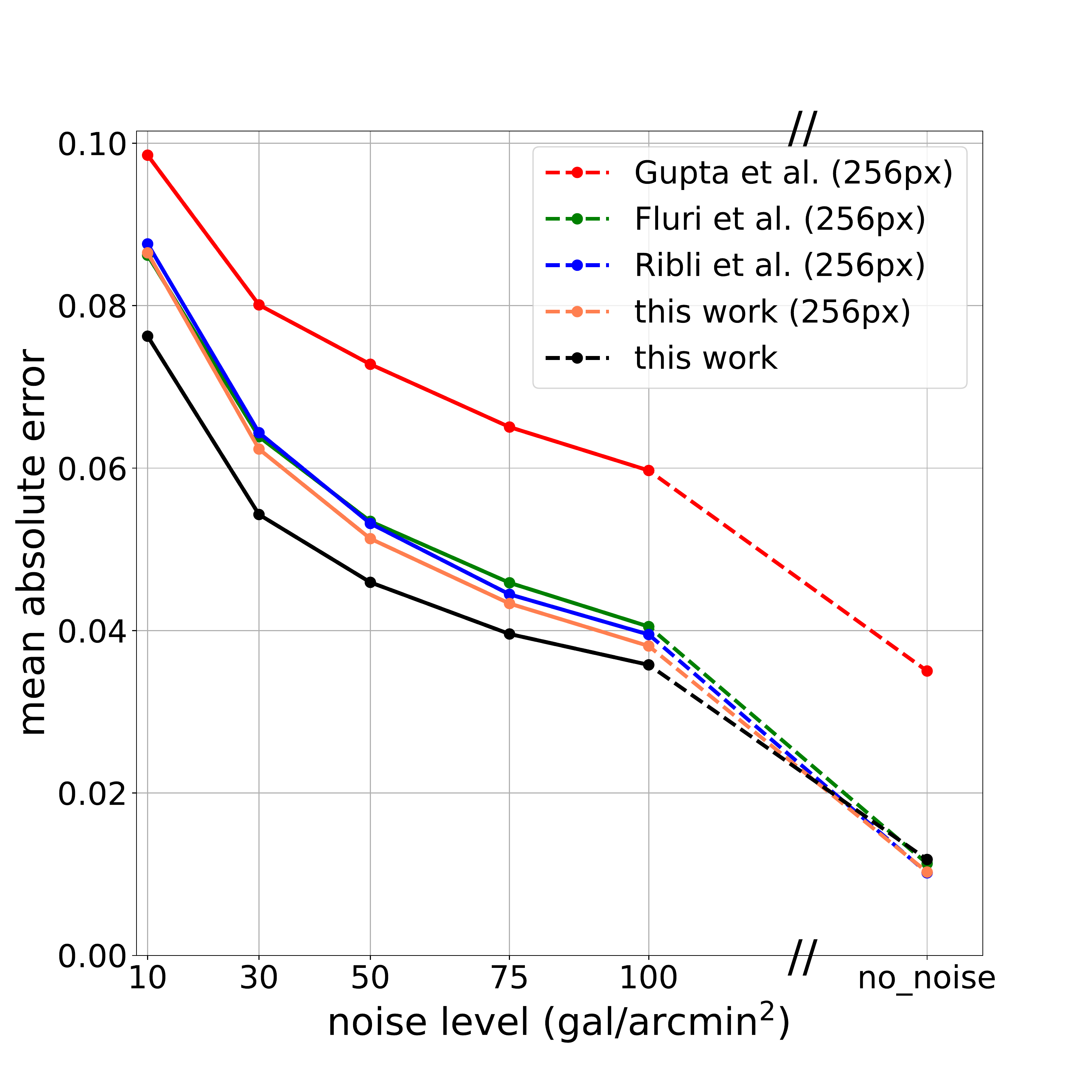}
    \caption{
    The CNN processing the whole convergence maps covering the $ 3.5 \times 3.5 \deg^2$  simulated field is more accurate, in the presence of shape noise, than the ones used in previous studies which only process smaller tiles.
    The advantage of using the full larger images is diminished on noiseless maps.  Specifically, different CNNs are trained at noiseless, 10, 30, 50, 75 and 100 galaxy arcmin$^{-2}$ shape noise levels. The vertical axis shows the mean absolute difference between the predicted and the simulated $\sigma_8$, $\Omega_m$ values. 
    CNNs used in previous studies cannot accept 512$\times$512-pixel maps, therefore maps were split to four 256$\times$256-pixel tiles for these networks, as indicated in the legend.
    The black line shows a CNN trained on 512$\times$512-pixel images.}
    \label{fig:shape_compare}
\end{figure}

Throughout this work we use varying and 'semi-varying' covariances, while some previous studies used fix covariances for some or all of the descriptors \citep{gupta2018non, fluri2018cosmological}. 
To understand the influence of this choice, we replicated our confidence contour results for each method using fixed covariance matrices.
We find that the power spectrum contours significantly improve when using varying covariances instead of fix ones.
On the other hand we find that the contours of the CNN {\em degrade} with varying covariances, possibly because the use of varying covariances correctly handle the effect of strong convergence of predictions on the covariance of predictions [Fig. \ref{fig:varying_vs_fixed}].
The results suggest that for a correct comparison of the CNN and other descriptors in this problem one needs to use varying or 'semi varying' covariances with each estimator. 
More details of the comparison are described in Appendix~\ref{subsec:varyingcov}.

In the main result section we adopted mock observations with parameters picked from the training set, therefore we did not force the CNN to interpolate between the points of the grid.
In a setup where the mock observation is not in the training dataset we can test the interpolation capabilities of the CNN.
After excluding the mock observation convergence maps from the training set we find that the predictions of the CNN do not degrade significantly, indicating that the neural network is able to interpolate the cosmological parameter grid [Fig. \ref{fig:unseen_params}].  Further details are given in Appendix~\ref{subsec:ninterpol}.

Each simulation in our training data also used the same random seed for the initial density and velocity fields, which results in very similar-looking convergence maps even with different cosmological parameters, when the maps are ray-traced from the same viewpoint.  
This raises the possibility that the network can memorize and make unfair use of the random phases in the initial conditions.
Therefore we next test the CNN on new convergence maps, simulated using a different random seed for the generation of initial density and velocity fields.
We find that the neural network is able to generalize to these fully independent new mock observations [Fig. \ref{fig:unseen_idf}]. Details are provided in Appendix~\ref{subsec:newidf}.

Augmenting the simulation suite by random transpositions and rotations are important, and helps mitigate overfitting. As discussed in Appendix~\ref{app:augmentation}, without this effective increase in the size of the training set, the network's performance would degrade significantly.

Another important issue is whether the views between the training and test sets are mutually exclusive. The impact of the method to split views between the training and test sets is discussed in Appendix~\ref{app:splits}. 
We find that mixing the same views in the training and test sets would lead to overestimating the accuracy of the CNN in the present paper, but does not make a difference in the less accurate CNNs used in our previous work.

Finally, the non-uniformity of the sampling of the cosmological parameter grid around the fiducial cosmology is investigated in Appendix~\ref{subsec:densegrid}.  We find that in general, uniform sampling is advantageous.

\subsection{Conclusions}
In this paper, we trained a convolutional neural network to predict the true underlying ($\Omega_m, \sigma_8$) cosmological parameters of simulated weak lensing convergence maps in the presence of shape noise levels 
corresponding to on-going and future large weak lensing surveys.
We show that the CNN is able to reduce credible parameter contour areas compared to the power spectrum by a factor or 3.5-4.2 for a deep space-based survey such as WFIRST and by a factor of 2.4-2.8 for LSST or Euclid.
We also show that the CNN is 1.4-2.1 more accurate than peak counts for a deep space-based survey such as WFIRST and by a factor of 1.05-1.42 for LSST or Euclid.
Our CNN is able to generalize to new, unseen initial density and velocity field realizations, and is capable of interpolating on the simulation grid.  These results indicate that cosmological parameter inference with convolutional neural networks could provide a large improvement for future large weak lensing surveys.   
Our results also suggest that the relative performance of a neural network degrades with increasing noise levels, and therefore implies that mitigation of shape noise is even more crucial than previously thought.

\section*{Acknowledgements}

This work was partially supported by National Research, Development and Innovation Office of Hungary via grant OTKA NN 129148 and the National Quantum Technologies Program. 
ZH acknowledges support from NASA ATP grant 
80NSSC18K1093.



\bibliographystyle{mnras}


\appendix

\section{Different neural network architectures and input map sizes}
\label{subsec:archcomp}

In order to understand the significance of using large inputs {\it vs.} the tiling scheme, and the effect of the neural network architecture used in this study, we retrain the networks presented in the previous studies \citep{gupta2018non, ribli2018learning, fluri2018cosmological}.
Those CNNs were designed for 256-pixel maps, therefore we retrain them on the four 256-pixel sized tiles (i.e. four quadrants) of our 512-pixel maps.
During inference, the predictions on single 512-pixel maps are the averages of the predictions on its four 256-pixel tiles.
We also retrain a variant of the network used in this present study on the 256-pixel sized tiles to separate the effect of the tiling scheme and the neural network details.
Our network needs to be modified in order to make it accept 256-pixel inputs by simply erasing a few layers.

The network presented in this study is trained with stochastic gradient descent optimizer for 30 epochs with an initial learning rate of $5\times10^{-3}$ dropped by a factor of 10 after the 10th, 15th, 20th and 25th epochs for stable convergence.
We have found that the networks used in \citep{gupta2018non, ribli2018learning, fluri2018cosmological} are unable to converge using the above-mentioned learning rate schedule, possibly due to the lack of batch-normalizations.
Therefore they are trained with an Adam optimizer for 30 epochs, with an initial learning rate of $5\times10^{-4}$ that was dropped by a factor of 10 after the 10th, 15th, 20th, and 25th epochs. 
We show the mean absolute error of predictions, depending on the level of noise for each network in Fig.~\ref{fig:shape_compare}.

We find that the neural networks used by \cite{ribli2018learning} and  \cite{fluri2018cosmological} perform very similarly on 256-pixel sized tiles as the variant of the network used in this study, although they are slightly less accurate.   
We also find in the presence of shape noise, feeding the CNN the 512-pixel convergence maps covering the whole $ 3.5 \times 3.5 \deg^2$  simulated field results in significantly more accurate predictions than the tiling scheme.

The results suggest that with relatively small simulated convergence maps the best strategy is to design a neural network capable of processing the full map instead of tiling to smaller regions.
With larger simulation boxes eventually one will probably have to cut the simulated convergence maps into smaller tiles, possibly due to GPU performance and memory limitations, however, the optimal input map size will have to be experimentally determined.

\section{The impact of cosmology-dependent covariances}
\label{subsec:varyingcov}

 \begin{figure}
   \includegraphics[width=\columnwidth]{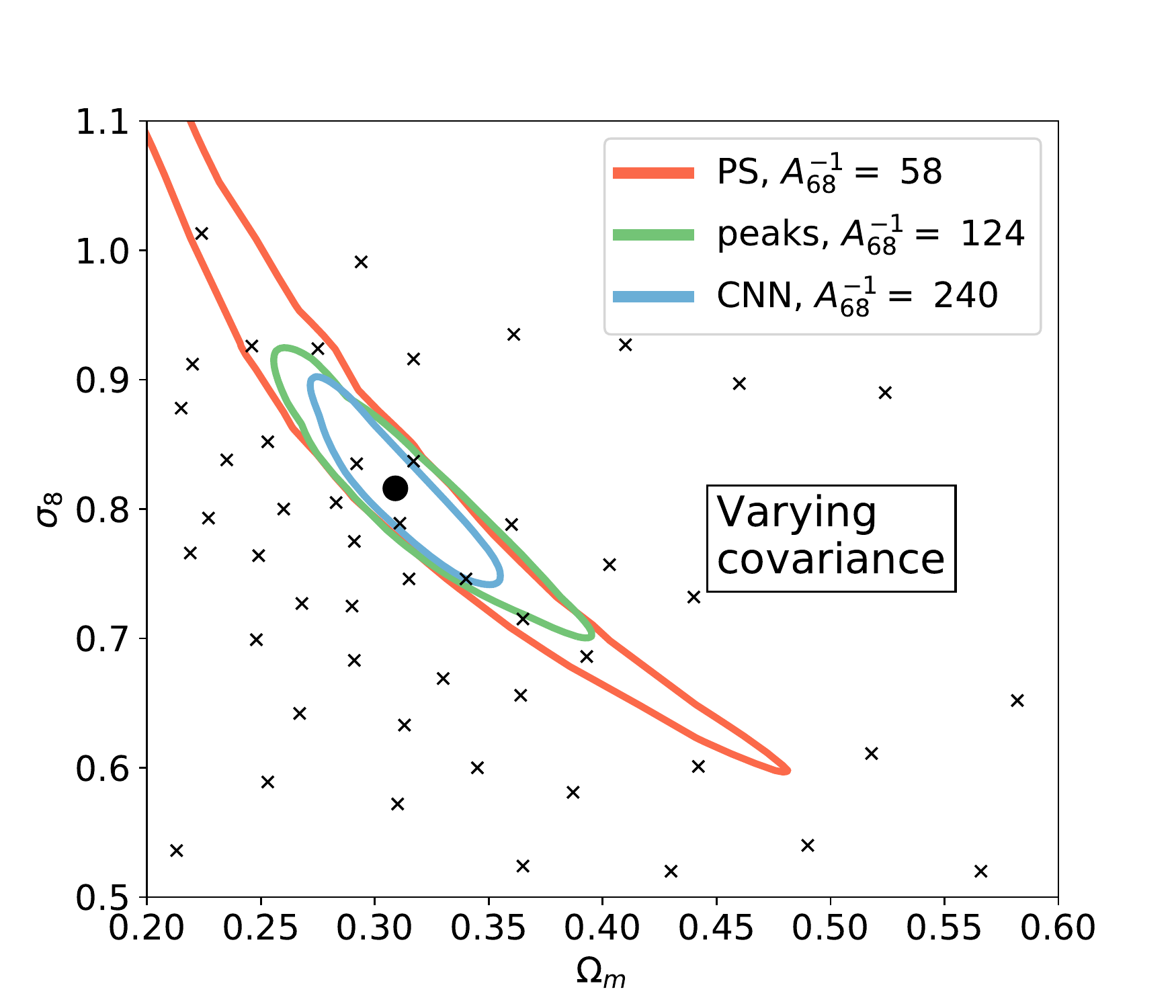}
    \includegraphics[width=\columnwidth]{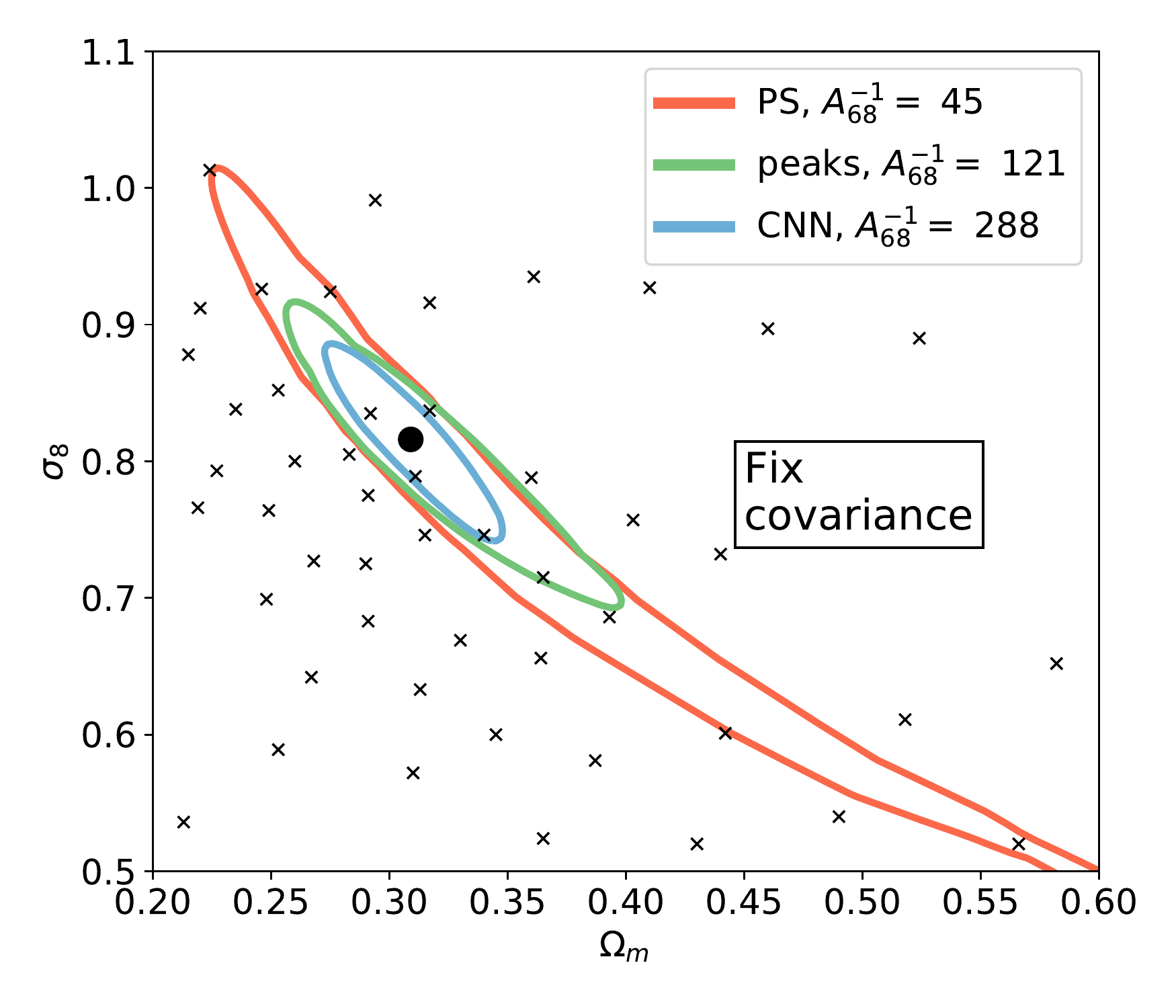}
    \caption{ Comparison of credible cosmological parameter contours using varying or semi-varying (Top)  and fixed covariances (Bottom) for a mock observation with $(\Omega_m=0.309, \sigma_8=0.816)$.
    Results are shown with shape noise from 75 galaxies arcmin$^{-2}$.
    The black crosses represent the parameter pairs used in simulations and the black dot marks the fiducial cosmological parameters.
    The panel shows the same results as [Fig \ref{fig:contours}] at only one noise level.}
    \label{fig:varying_vs_fixed}
\end{figure}

In this work we use semi-varying covariances for peak counts and the power spectrum, and varying covariance for the CNN, while previous studies used fixed covariances evaluated at the fiducial cosmological parameters for each method \citep{gupta2018non} or for the power spectrum alone \citep{fluri2018cosmological}.
In order to understand the effects of fixed {\it vs.} varying covariances on our results, we re-evaluate the confidence contours with fixed covariances at a mock observation with ($\Omega_m=0.309, \sigma_8=0.816$).  The results are shown in Fig.~\ref{fig:varying_vs_fixed}.

In agreement with a previous study \citep{fluri2018cosmological} we find that the contours for the CNN are slightly larger when using a varying covariance instead of a fixed one. 
Similarly to another previous study \citep{matilla2016dark} we also find that the confidence contours derived with peak counts are smaller when evaluating the likelihoods with semi-varying covariances.
Interestingly, we find that the contours of the power spectrum are also around 1.3 times smaller when using semi-varying covariances compared to fixed ones.
This result indicates that in order to fairly compare constraints with the CNN and the power spectrum, one needs to also evaluate the power spectrum with varying covariances, not only the CNN.

Unexpectedly (at least to us), allowing covariances to depend on cosmology has the opposite effect on the constraints derived from the CNN (tightening constraints) and on the pre-specified statistics (degrading the constraints).
The main reason for the different direction of the effect is that the predictions of the CNN are converging towards the center of our parameter grid and therefore towards the fiducial parameters [Fig. \ref{fig:bias}].
This apparent convergence results in artificially small covariances at the fiducial parameters compared to other grid points.
Therefore using a fixed covariance evaluated at the fiducial cosmology underestimates the size of the credible confidence contours.

\section{Cosmological parameter interpolation}
\label{subsec:ninterpol}

Testing the predictions of the neural network using maps with cosmological parameters which are included in the training set does not require the CNN to interpolate between the points on the simulated parameter grid.
In order to test the capability of the network to interpolate cosmological parameters, we retrain the neural network after excluding the fiducial parameters  $(\Omega_m=0.309, \sigma_8=0.816$) from the training set.
Therefore during testing the CNN on maps with fiducial parameters we force the network to interpolate the learned cosmological parameters on the grid.

\begin{figure}
    \includegraphics[width=\columnwidth]{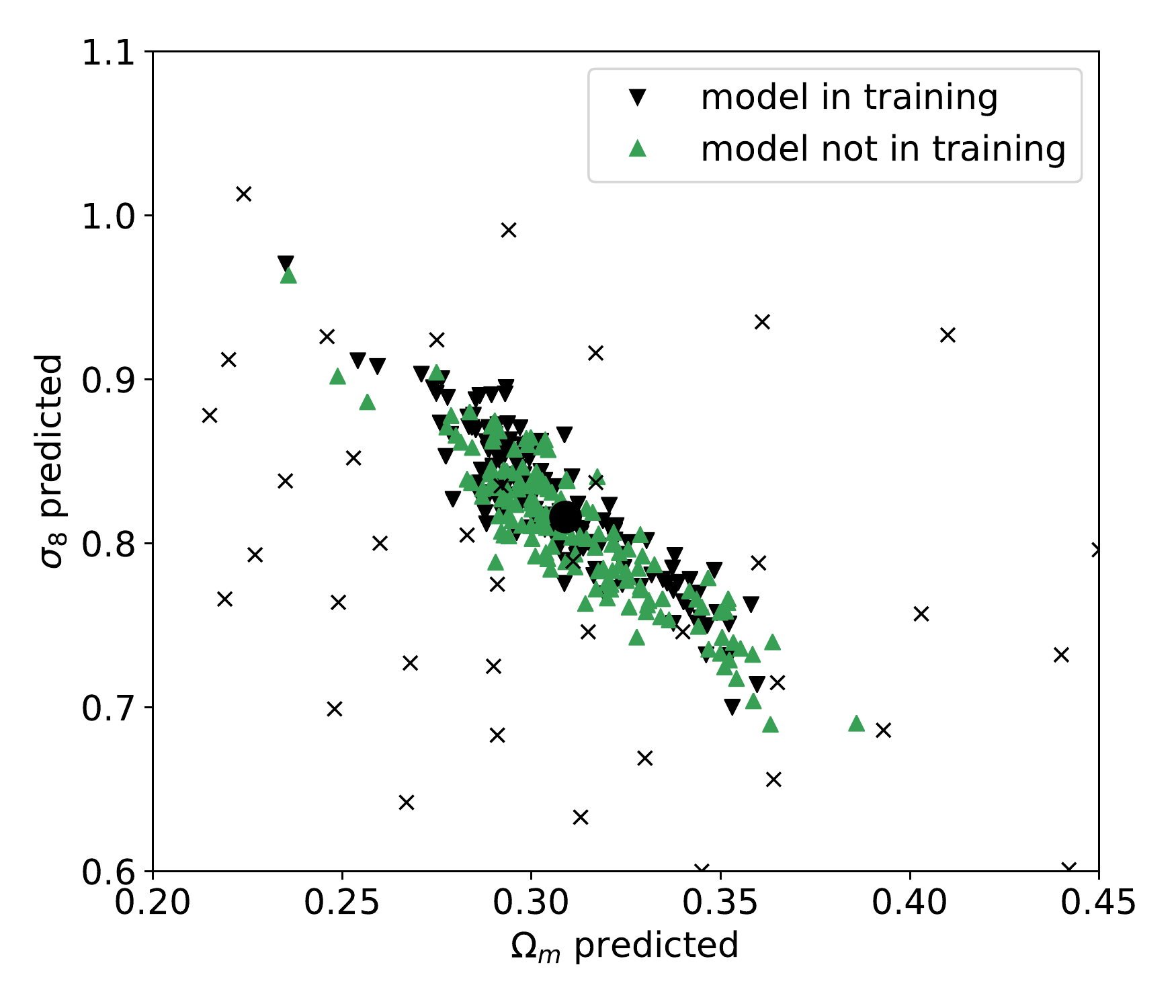}
    \includegraphics[width=\columnwidth]{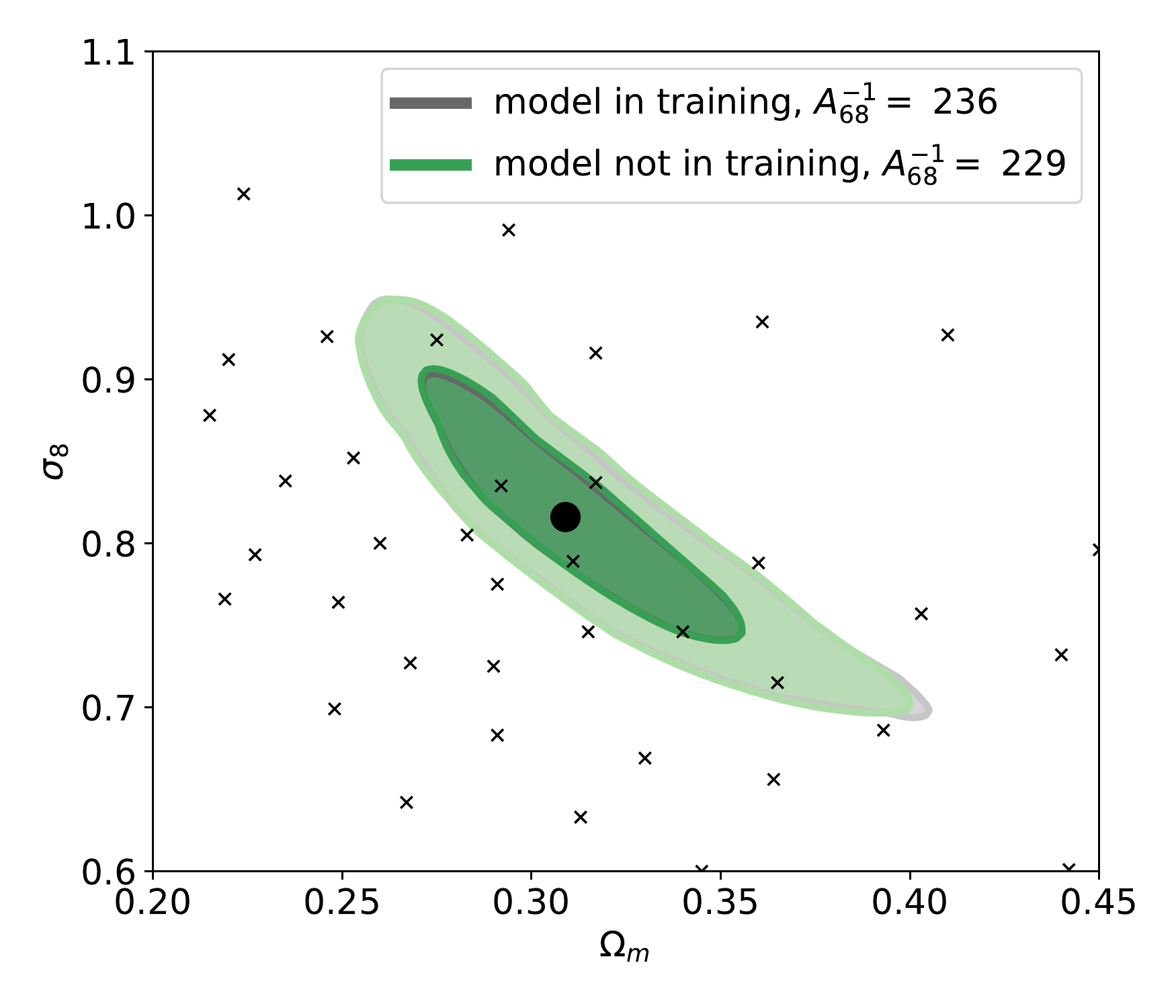}
    \caption{
    The CNN is able to interpolate the cosmological parameters of convergence maps. Predictions of the CNN and credible parameter contours do not change radically if we exclude the fiducial cosmological parameters from the training set.
    \textbf{Top}: Predictions on lensing maps with fiducial cosmological parameters ($\Omega_m=0.309, \sigma_8=0.816$), before and after excluding the maps with fiducial parameters from the training set.
    \textbf{Bottom}: Credible contours for the fiducial mock observation ($\Omega_m=0.309, \sigma_8=0.816$)  before and after excluding the maps with fiducial parameters from the training set.
    Noise level of 75 galaxies per square-arcmin.
    The black crosses represent the parameter pairs used in simulations, the black dot marks the fiducial cosmological parameters.}
    \label{fig:unseen_params}
\end{figure}

The predictions on the maps with fiducial parameters have similar distributions regardless of whether the fiducial parameters were included or excluded from the training set [Fig. \ref{fig:unseen_params}].
The credible contours derived after excluding the fiducial parameter from the training set are not significantly different compared to the original ones, and their area is essentially the same [Fig. \ref{fig:unseen_params}].
The results indicate that the neural network is indeed capable of interpolation between the original points of the cosmological parameter grid.

\section{Initial density and velocity fields generated with a different random seed}
\label{subsec:newidf}

\begin{figure}
	\includegraphics[width=\columnwidth]{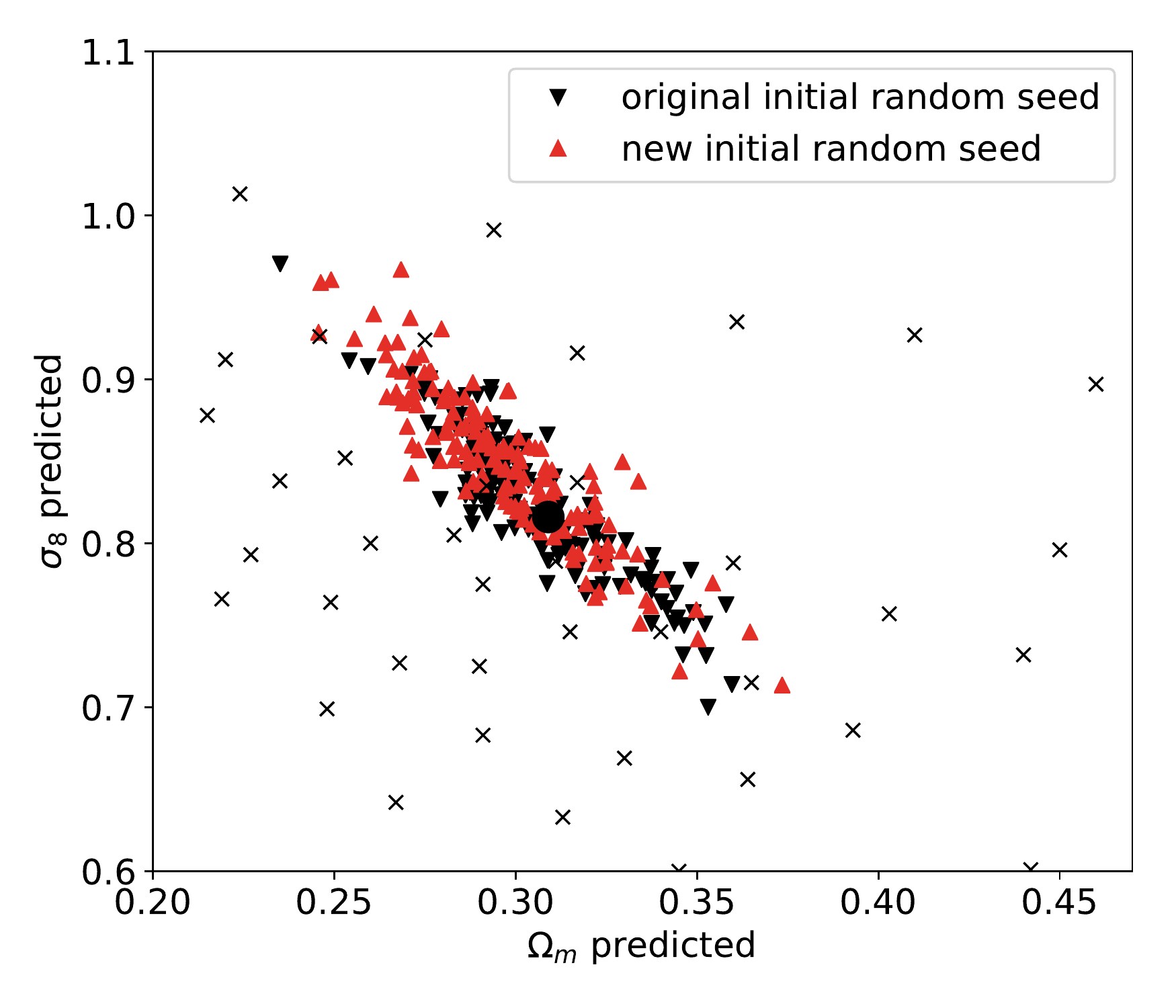}
	\includegraphics[width=\columnwidth]{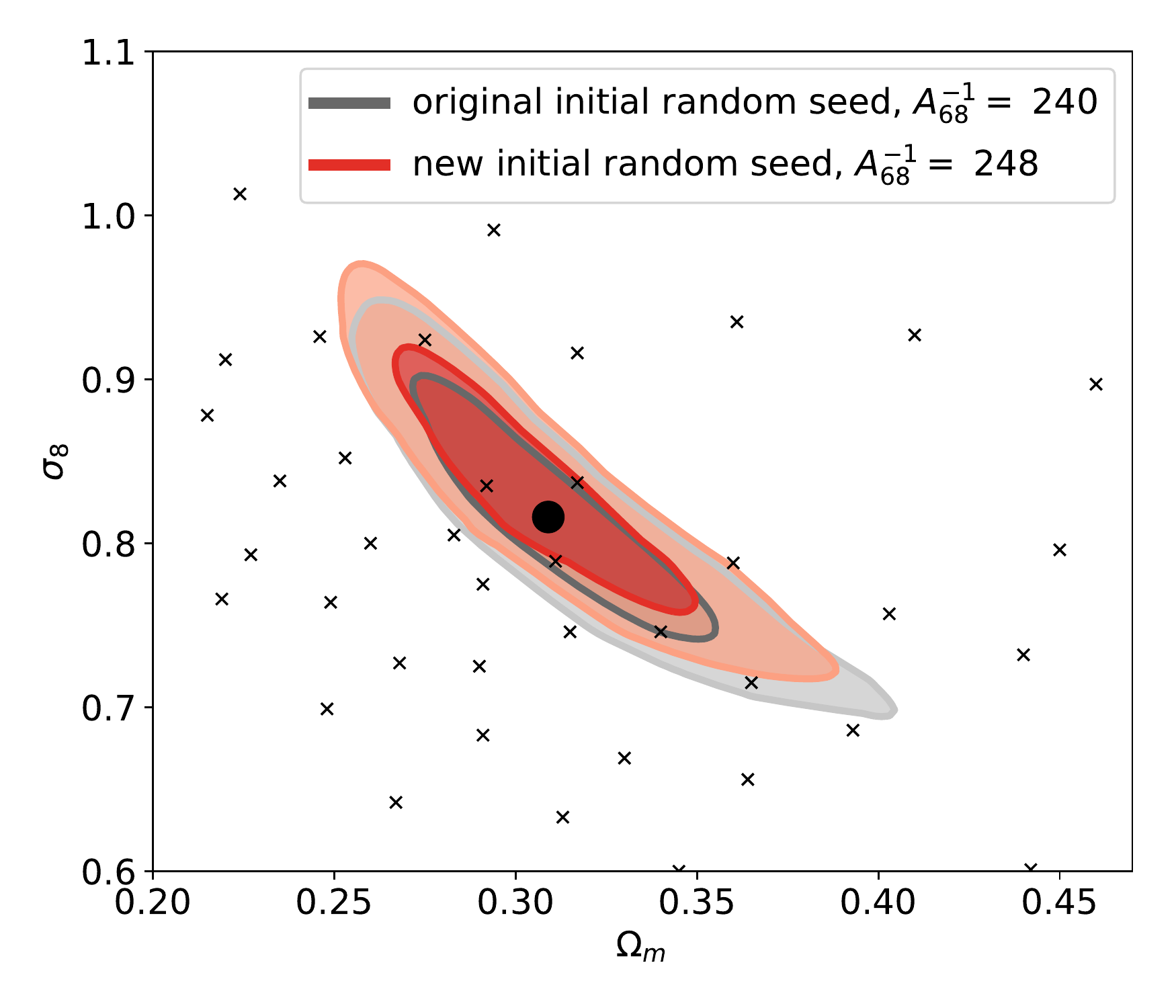}
    \caption{
    The CNN is able to generalize to new initial density and velocity fields.
    Predictions of the CNN and credible parameter contours do not change radically if a different random seed is used for the generation of the initial density and velocity field.
    \textbf{Top}: Predictions on lensing maps generated with the original and the new initial random seed with underlying true cosmological parameters ($\Omega_m=0.309, \sigma_8=0.816$).
    \textbf{Bottom}: Credible contours for a mock observation ($\Omega_m=0.309, \sigma_8=0.816$) before and after replacing the lensing maps at the mock observation to maps generated with a new initial random seed.
    Noise level of 75 galaxies arcmin$^{-2}$.
    The black crosses represent the parameter pairs used in simulations and the black dot marks the fiducial cosmological parameters.}
    \label{fig:unseen_idf}
\end{figure}

Our $N$-body simulations use initial density and velocity fields generated with the same random seed.
The original motivation for this choice, in the context of the pre-specified statistics (power spectrum and peak counts) was to measure the cosmology-dependence of their expectation value, uncontaminated by small random noise.   Likewise, using the same random seed also helps the neural network learn cosmology instead of the variation in initial conditions.
Randomized recycling (i.e. random rotations and translations of the N-body boxes) is then used to generate pseudo-independent realizations which were shown to be independent in terms of their power spectrum and peak counts \citep{petri2016sample}.
However, as shown previously, a convolutional neural network is able to extract more information from convergence maps than these descriptors.
In order to test the effect of a different initial density and velocity fields on our results, we evaluated the CNN using lensing maps from a simulation with $\Omega_m=0.309, \sigma_8=0.816$, where the initial density and velocity fields were generated using a different random seed.

The predictions from these new maps, shown in Fig.~\ref{fig:unseen_idf}, have a very similar distribution to the original versions.  In particular, the new credible contours are not significantly biased compared to the original ones, and their area remains essentially the same. The results indicate that the CNN is able to generalize to new initial density and velocity fields generated with a different random seed.

\section{The importance of data augmentation}
\label{app:augmentation}
\begin{figure}
  \includegraphics[width=1.0\columnwidth]{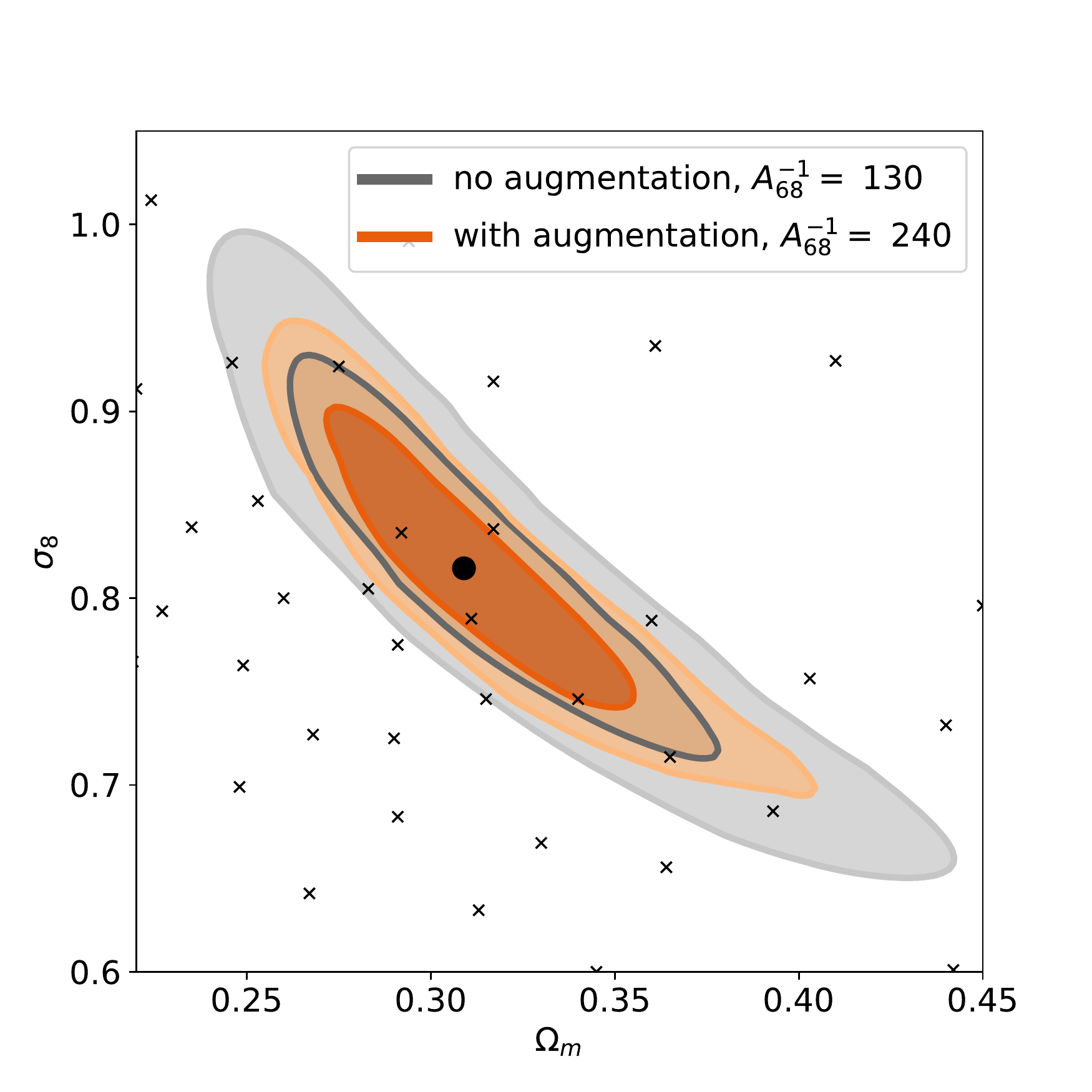} 
  \caption{Data augmentation is essential for the dataset used in this work.
  Credible parameter contours are 50\% smaller with data augmentation.
  Results are shown with shape noise corresponding to 75 galaxies arcmin$^{-2}$.}
  \label{fig:noaug}
 \end{figure}

During the training we process maps with random horizontal and vertical flips and random transpositions in order to further enrich the dataset.
To evaluate the significance of this data augmentation, we retrained the CNN without any data augmentation either during training or testing.
We find that without augmentation, the training error is significantly reduced, but the test error is increased. This shows that without augmentation, the neural network strongly overfits the training data [Fig. \ref{fig:noaug}], and this overfitting reduces its ability to correctly predict unseen maps from test set.
The inferred credible contours show a 50\% reduction in contour size due to data augmentation, suggesting that augmentation is indeed essential to obtain a good result with the current dataset.
Note that this result is partly due to the fact that each simulation is generated using the same random seed for the initial density and velocity field, further discussed in the next section.

\section{Training and test split using recycled simulations with the same initial seed}
\label{app:splits}

 \begin{figure}
	\includegraphics[width=0.9\columnwidth]{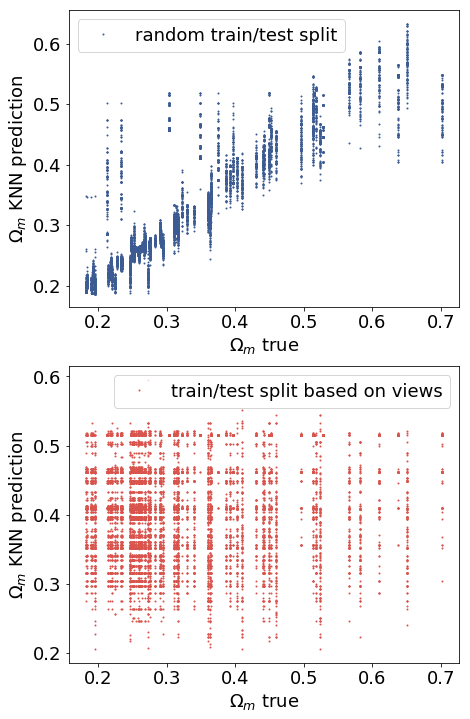} 
  \caption{Top: KNN regressor in pixel space predicts $\Omega_m$ very accurately if the training and the testing data is split randomly.
  Bottom: If the split is based on viewpoints a KNN in pixel space predicts $\Omega_m$ randomly as expected.}
  \label{fig:knn}
\end{figure}

\begin{figure}
  \includegraphics[width=\columnwidth]{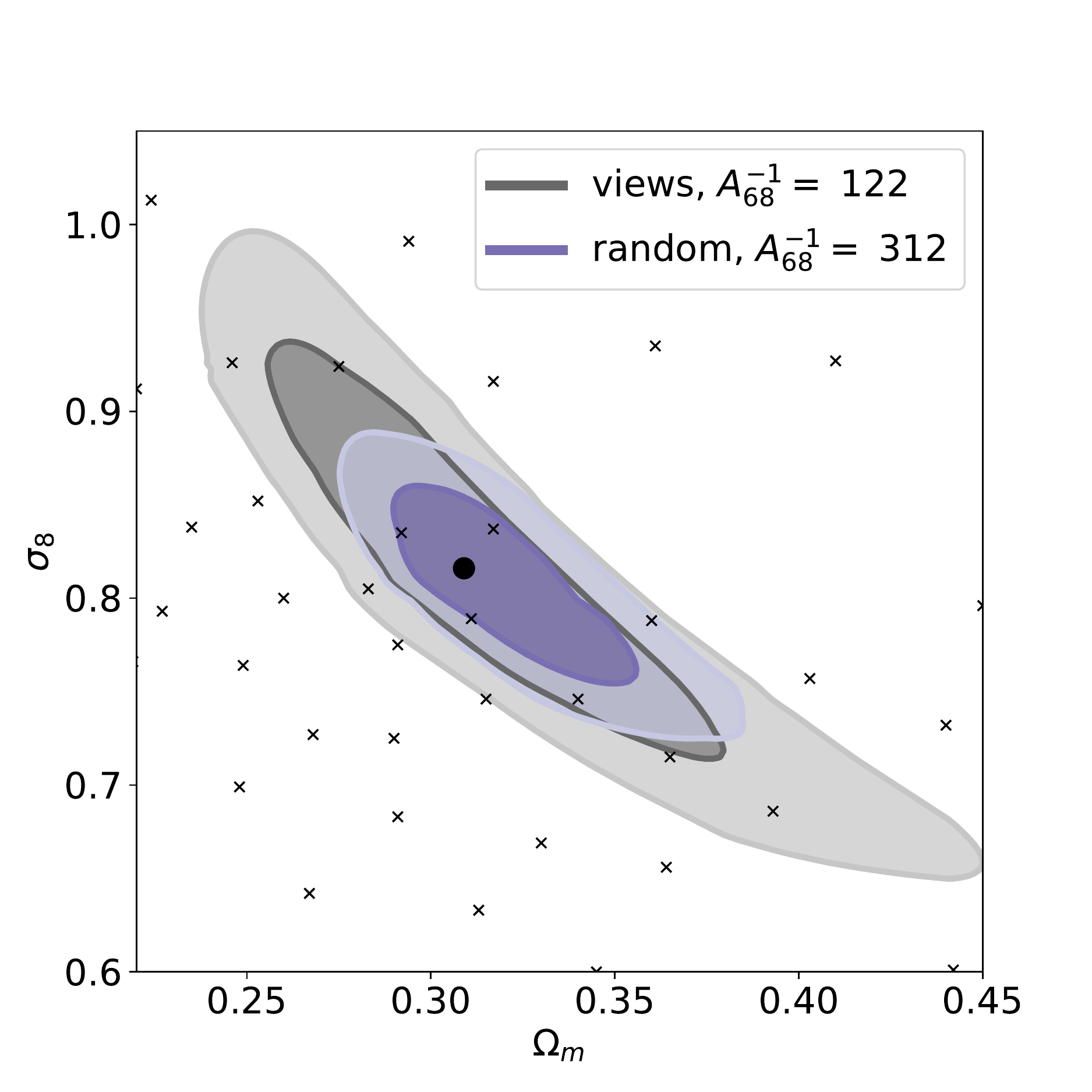} 
  \caption{The neural network yields spuriously smaller confidence contours when evaluated with a random train-test split instead of the split based on viewpoints for ray-tracing. Noise level: 75 galaxies arcmin$^{-2}$. No data augmentation used.}
  \label{fig:randomsplits}
\end{figure}

Large convolutional neural networks have very high capacity, allowing them to simply memorize the training data, even if it is 1.2 million images with completely random labels or pixels, as in \cite{zhang2016understanding}.
Therefore one needs to be very careful to test neural networks on examples as close to a real application setup as possible in order to test their true generalization capabilities.
In the dataset used for the study, the convergence maps are created with initial density and velocity fields generated with the same random seed, and pseudo-independent realizations from a simulation are generated from different viewpoints.
Simulations with the same initial conditions and viewpoints and slightly different parameters yield $\kappa$ maps that look almost identical by eye.
If we randomly assign the pseudo-independent realizations into a training and a testing set, as in previous studies \citep{gupta2018non,ribli2018learning}, we might be overestimating the accuracy of our neural network.

Let's consider an example: \textbf{A} map in the training set comes from view \#$1$ with parameters $(0.26,0.8)$, and \textbf{B} map in the test set comes from the same view \#$1$ but it has parameters $(0.268,0.81)$.
A and B maps are almost identical because they come from simulations with the same initial conditions, and only very slightly different cosmological parameters, and they are ray traced using the same past light cone.
Therefore if the neural network memorizes that \textbf{A} map has parameters $(0.26,0.8)$, and predicts that \textbf{B} map has the same parameters, it will only make a very small error, even though it has not learned a single thing about cosmology.

Assuming that the different realizations of a simulation are truly independent, a correct testing setup is easy to construct: maps need to be consistently split into the training and the testing set based on their viewpoints.
One past light cone for ray-tracing can only appear in either the test set or the training set, regardless of the cosmological parameters used to create the map.

In order to demonstrate the effect of splitting based on point of views, we create an experiment with a simple toy model, which is not able to learn information about cosmology, but it is very effective at memorizing maps: a k-nearest-neighbor regressor (KNN) with raw pixel inputs and Euclidean metric, with 4 neighbors, on low resolution (downsampled to 6 arcmins per pixel) convergence maps. 
For this test, we use the scikit-learn implementation of KNN.
If we split the maps randomly, the accuracy of the KNN is artificially very high, but if we split based on the viewpoints the high accuracy completely diminishes, as expected for such a simple model operating in the raw pixel space [Fig. \ref{fig:knn}].

\begin{figure}
	\includegraphics[width=\columnwidth]{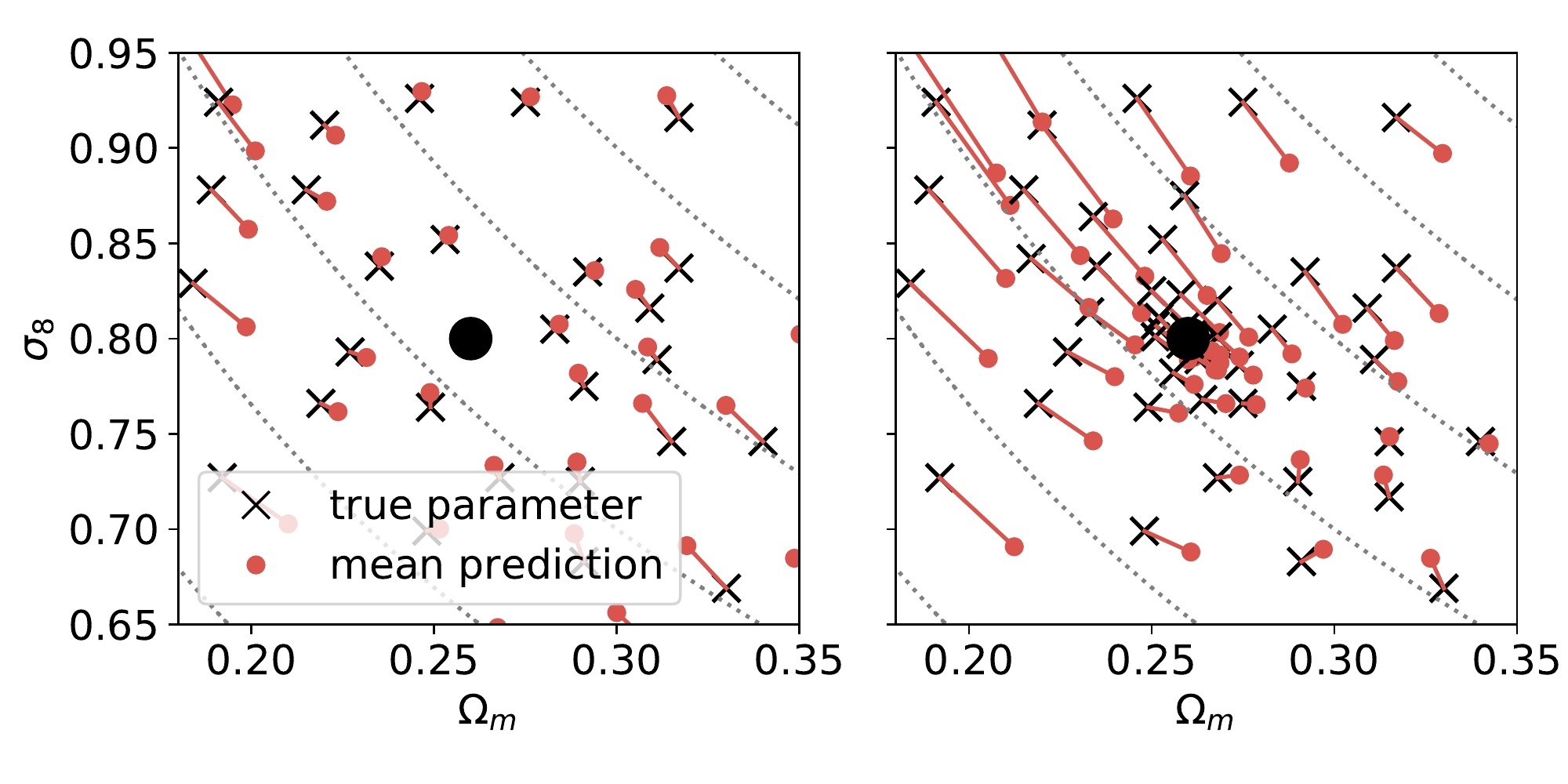} 
    \caption{Additional grid points around a cosmology create convergent predictions towards the densely sampled region.
    The left panel shows mean predictions with the quasi-uniformly sampled grid used in this study, the right panel shows mean predictions with additional grid points around cosmology $(\Omega_m=0.26, \sigma_8 = 0.8 ) $.
    The black x represents the true cosmological parameters used in the simulations, the red dot shows the mean of the predictions on the test maps, and the red line depicts the bias of the predictions.}
    \label{fig:densegrid_conv}
    \includegraphics[width=\columnwidth]{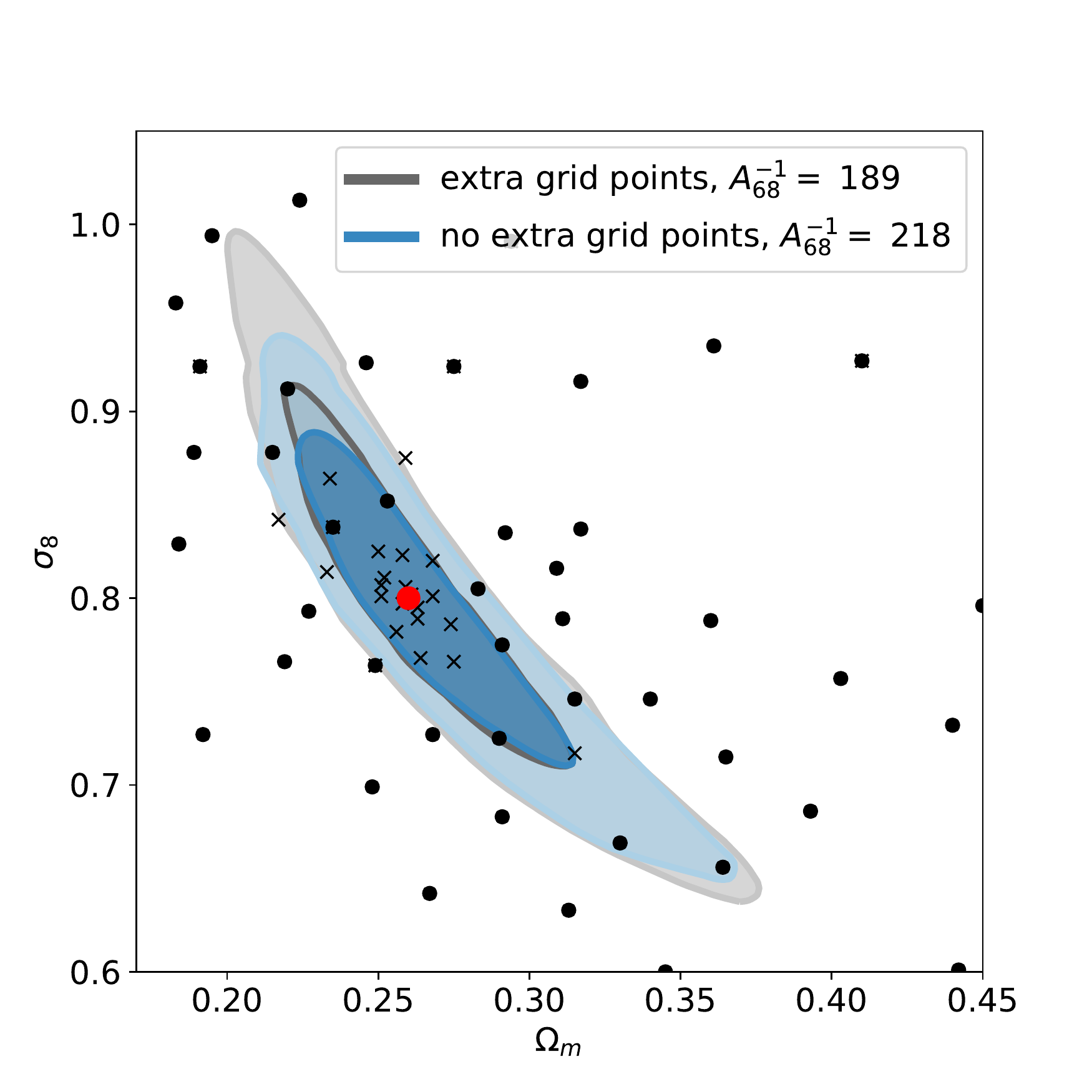} 
    \caption{Additional grid points around a cosmology surprisingly degrade contours instead of improving them if the mock observation falls in the more densely sampled region.
      Original grid points are marked with black circles. Additional grid points are marked with black crosses, and cosmology $(\Omega_m=0.26, \sigma_8 = 0.8 ) $ is marked with a red circle.}
      \label{fig:densegrid}
 \end{figure}

We also retrained and evaluated the neural network on the original split based on views, and a random train-test split, without data augmentations and L2 regularization in order to allow the network to potentially overfit the data.
In the case of the random splits the predictions of the network are artificially more accurate on the test set and the contours are significantly smaller, compared to the split based on viewpoints, demonstrating spurious improvements due to the incorrect testing setup [Fig.  \ref{fig:randomsplits}].
Here we find that this improvement diminishes when we use data augmentation and L2 regularization which reduce the problem of augmentation, however, there is no guarantee that the issue does not return with a different model or after other changes. 
The only safe way to overcome the information leak is to use properly split training and test data.

The issue with random splits was not recognized in previous studies \citep{gupta2018non, ribli2018learning}, therefore we repeated the experiments of those papers in order to evaluate the significance of the splits.
We find that the results of those studies do not change significantly if we split by views instead of a random split originally used in these studies, possibly due to the fact that those studies used some form of data augmentation and they trained the networks for a very few iterations, only 5 epochs.
Early stopping during training is regarded as an effective way of regularization \citep{Goodfellow}.

\section{Additional grid points around one cosmology}
\label{subsec:densegrid}

In the present study, we used a quasi-uniformly sampled cosmological parameter grid, without increasing the sampling density in regions around the fiducial cosmology as was done in previous work \cite{gupta2018non}.
We conduct an experiment in order to evaluate whether additional points only around one cosmology $(\Omega_m=0.26, \sigma_8 = 0.8)$ as in \cite{gupta2018non} improve the predictions.
We find that additional points around one cosmology create a stronger artificial convergence of the predictions towards the more densely sampled region [Fig. \ref{fig:densegrid_conv}].
After the inversion of raw predictions into credible confidence contours, the convergence due to the additional grid points {\em degrades} the quality of credible parameter contours instead of improving them [Fig. \ref{fig:densegrid}].
The result suggests that the density of the simulation grid should be quasi-uniform in order to achieve the best results.

\begin{figure}   \includegraphics[width=0.9\columnwidth]{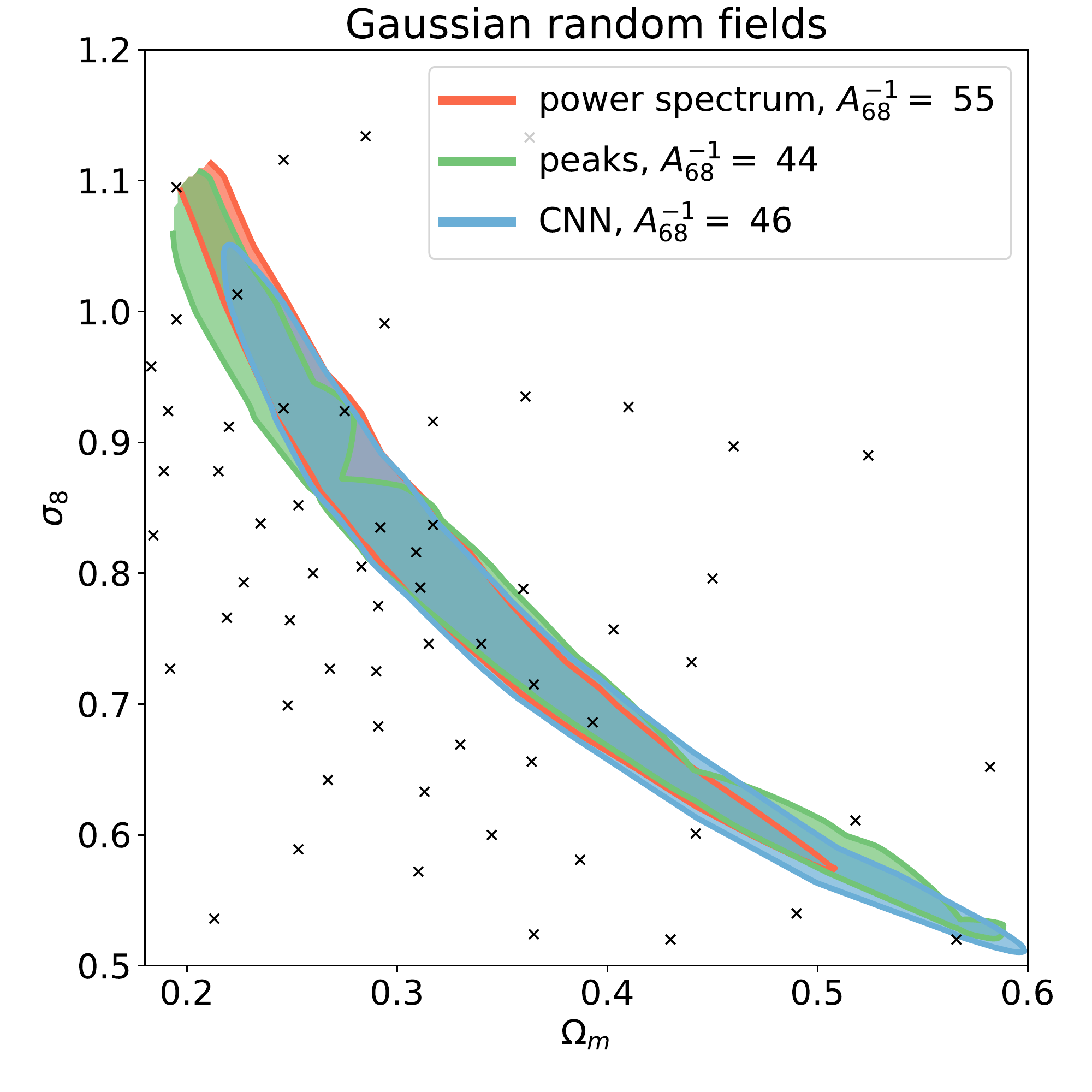}
    \caption{Neither the CNN nor peak counts are able to extract more information from Gaussian random fields than the power spectrum. The figure shows results from a shape noise level of 75 galaxies arcmin$^{-2}$.}
    \label{fig:grf_test}
\end{figure}

\section{Results on equivalent Gaussian random fields}

Finally, we performed a null-test similar to \cite{gupta2018non}: we  examined whether the neural network or peak counts outperforms the power spectrum when using Gaussian random fields (GRF) as inputs instead of the physical convergence maps.  If it did, it would indicate over-fitting.
We found that on GRF inputs the power spectrum is the most accurate as expected [Fig. \ref{fig:grf_test}], i.e. the network passed this test.

\bsp	
\label{lastpage}
\end{document}